\documentclass[preprint2]{aastex631}

\usepackage{graphicx}	
\usepackage{amsmath}

\begin{document}

\title{``Oh FUors where art thou'': A search for long-lasting YSO outbursts hiding in infrared surveys.}
\shorttitle{The search for FUor outbursts}

\author[0000-0003-1894-1880]{Carlos Contreras Pe\~{n}a}
\affiliation{Department of Physics and Astronomy, Seoul National University, 1 Gwanak-ro, Gwanak-gu, Seoul 08826, Korea}
\affiliation{Research Institute of Basic Sciences, Seoul National University, Seoul 08826, Republic of Korea}

\author[0000-0003-3119-2087]{Jeong-Eun Lee}
\affiliation{Department of Physics and Astronomy, Seoul National University, 1 Gwanak-ro, Gwanak-gu, Seoul 08826, Korea}
\affiliation{SNU Astronomy Research Center, Seoul National University, 1 Gwanak-ro, Gwanak-gu, Seoul 08826, Korea}

\author[0000-0002-3808-7143]{Ho-Gyu Lee}
\affiliation{Korea Astronomy and Space Science Institute, 776 Daedeok-daero, Yuseong, Daejeon 34055, Korea}

\author[0000-0002-7154-6065]{Gregory Herczeg}
\affiliation{Kavli Institute for Astronomy and Astrophysics, Peking University, Yiheyuan Lu 5, Haidian Qu, 100871 Beijing, Peoples Republic of China}
\affiliation{Department of Astronomy, Peking University, Yiheyuan 5, Haidian Qu, 100871 Beijing, China}

\author[0000-0002-6773-459X]{Doug Johnstone}
\affiliation{NRC Herzberg Astronomy and Astrophysics, 5071 West Saanich Rd, Victoria, BC, V9E 2E7, Canada}
\affiliation{Department of Physics and Astronomy, University of Victoria, Victoria, BC, V8P 5C2, Canada}

\author{Hanpu Liu}
\affiliation{Kavli Institute for Astronomy and Astrophysics, Peking University, Yiheyuan Lu 5, Haidian Qu, 100871 Beijing, Peoples Republic of China}
\affiliation{Department of Astronomy, Peking University, Yiheyuan 5, Haidian Qu, 100871 Beijing, China}

\author{Philip W. Lucas}
\affiliation{Centre for Astrophysics Research, University of Hertfordshire, College Lane, Hatfield, AL10 9AB, UK}

\author[0000-0003-0292-4832]{Zhen Guo}
\affiliation{Instituto de Fisica y Astronomia, Universidad de Valparaiso, ave. Gran Breta\~{n}a, 1111, Casilla 5030, Valparaiso, Chile}
\affiliation{Millennium Institute of Astrophysics, Nuncio Monse{\~n}or Sotero Sanz 100, Of. 104, Providencia, Santiago, Chile}
\affiliation{Centre for Astrophysics Research, University of Hertfordshire, College Lane, Hatfield, AL10 9AB, UK}

\author[0000-0002-0631-7514]{Michael A. Kuhn}
\affiliation{Centre for Astrophysics Research, University of Hertfordshire, College Lane, Hatfield, AL10 9AB, UK}

\author{Leigh C. Smith}
\affiliation{Institute of Astronomy, University of Cambridge, Madingley Road, Cambridge CB3 0HA, UK}

\author{Mizna Ashraf}
\affiliation{Department of Physics, Indian Institute of Science Education and Research Tirupati, Yerpedu, Tirupati - 517619, Andhra Pradesh, India.}

\author{Jessy Jose}
\affiliation{Department of Physics, Indian Institute of Science Education and Research Tirupati, Yerpedu, Tirupati - 517619, Andhra Pradesh, India.}

\author{Sung-Yong Yoon}
\affiliation{Korea Astronomy and Space Science Institute, 776 Daedeok-daero, Yuseong, Daejeon 34055, Korea}
\affiliation{School of Space Research, Kyung Hee University, 1732, Deogyeong-daero, Giheung-gu, Yongin-si, Gyeonggi-do 17104, Republic of Korea}

\author{Sung-Chul Yoon}
\affiliation{Department of Physics and Astronomy, Seoul National University, 1 Gwanak-ro, Gwanak-gu, Seoul 08826, Korea}

\correspondingauthor{Carlos Contreras Pe\~{n}a}
\email{cecontrep@gmail.com, ccontreras@snu.ac.kr}



\begin{abstract}

Long-lasting episodes of high accretion can strongly impact stellar and planetary formation. However, the universality of these events during the formation of young stellar objects (YSOs) is still under debate. Accurate statistics of strong outbursts (FUors), are necessary to understand the role of episodic accretion bursts. In this work, we search for a population of FUors that may have gone undetected in the past because they either a) went into outburst before the start of modern monitoring surveys and are now slowly fading back into quiescence or b) are slow-rising outbursts that would not commonly be classified as candidate FUors. We hypothesise that the light curves of these outbursts should be well fitted by linear models with negative (declining) or positive (rising) slopes. The analysis of the infrared light curves and photometry of $\sim$99000 YSO candidates from SPICY yields 717 candidate FUors. Infrared spectroscopy of 20 candidates, from both the literature and obtained by our group, confirms that 18 YSOs are going through long-term outbursts and identifies two evolved sources as contaminants. The number of candidate FUors combined with previously measured values of the frequency of FUor outbursts, yield average outburst decay times that are 2.5 times longer than the rise times. In addition, a population of outbursts with rise timescales between 2000 and 5000 days must exist to obtain our observed number of YSOs with positive slopes. Finally, we estimate a mean burst lifetime of between 45 and 100 years.

\end{abstract}



\section{Introduction} \label{sec:intro}

In the episodic accretion model of star formation, stars gain most of their mass in short-lived episodes of high accretion followed by long periods of quiescent low-level accretion \citep[see e.g.][]{2009Enoch, 2014Bae,2014Dunham,2023Fischer}. Episodic accretion can have lasting effects on both stellar and planetary formation. It has been invoked as one of the solutions to the observed spread in the luminosities of Class I YSOs  \citep[e.g.][]{1990Kenyon,2009Evans}. The long time spent at high accretion rates can have long-term effects on the structure of the central star \citep[e.g.][]{2017Baraffe,2017Kunitomo}. The associated outbursts can alter the chemistry of protoplanetary discs and envelopes \citep{2007Lee_JE,2013Jorgensen,2024Zwicky}, move the location of the snowline of various ices \citep{2016Cieza, 2019jelee}, aid in the formation of planetary systems similar to the solar system \citep{2017Hubbarda}, and affect the orbital evolution of planets, if present \citep{2013Boss,2021Becker}. Despite all these possibilities, however, the role of episodic accretion in stellar mass assembly is still uncertain.

A class of young stellar objects (YSOs), known as eruptive variable YSOs (a.k.a.\ FUors/EX Lup-like), display high-amplitude variability due to a large increase in the accretion rate onto the central star, supporting the episodic accretion model. The most extreme of these outbursts in YSOs, known as FUors after the prototype FU Ori, have accretion rates that increase from normal T Tauri rates of 10$^{-8}$--10$^{-9}$~M$_{\odot}$ yr$^{-1}$ to reach as high as 10$^{-4}$ M$_{\odot}$ yr$^{-1}$, and last for centuries \citep{1996Hartmann}. FUors are detected and classified according to the photometric and spectroscopic characteristics of the outbursts.

The discovery of additional FUors typically requires the observation of a sudden and large amplitude outburst, similar to the brightness changes observed in known FUors, such as FU Ori \citep{1966Herbig,1977Herbig}, HBC722 \citep{2010Semkov} or the majority of newly discovered FUors from the VVV survey \citep{2024Guo_a}. This requirement, however, fails to allow for sources that are already undergoing high-accretion outbursts. Firstly, after reaching maximum brightness, FUor outbursts slowly decay over the next  $\sim$ 10 to 100 yr. For example, the brightness of FU Ori has been declining since its outburst in 1936 at an average rate of 0.015 mag yr$^{-1}$ \citep{2000Kenyon}. Thus, if the start of a monitoring campaign misses the outburst in a particular YSO, the slow declining light curve would not be identified as a potential candidate FUor. Secondly, long-term rising outbursts \citep[such as SSTgbs J21470601$+$4739394,][]{2024Ashraf} might not be immediately recognised as being driven by large changes in the accretion rate.  Therefore, there should exist a population of FUors that remain undetected by the standard discovery criterion.

The confirmation as a bona-fide FUor requires spectroscopic follow-up, as these objects display unique characteristics during the outburst \citep[][]{2019Hillenbrand_b,2022Rodriguez}. For example, the near-infrared spectrum of FUors shows a triangular shape H-band continuum due to H$_2$O absorption and strong first overtone $^{12}$CO absorption at 2.3--2.4 $\mu$m \citep[e.g.][]{2018Connelley}. Hydrogen emission lines, a common feature in YSOs, are usually weak or not present in the spectrum of FUors \citep{2010Connelley,2012Fischer}. These features are explained by a rapidly accreting disk that outshines the central protostar at all wavelengths. The absorption lines form in a disk that is heated from the midplane and has a cooler atmosphere \citep[e.g., ][]{1991Calvet,1996Hartmann}.

The unique spectroscopic characteristics of FUor outbursts have allowed for the classification of FUor-like sources \citep[see e.g.][]{1985Mundt,1997Reipurth_aspin,2010Connelley}. These YSOs show the same spectroscopic features as FUors, but where no outburst has been recorded. As FUors have short rise times compared with the expected lifetime in the bright state, outbursts in FUor-like sources likely would have occurred before the start of active monitoring of star-forming regions (SFRs).
FUor-like sources are often discovered through spectroscopic studies of stars with particularly interesting properties \citep[such as sources driving Herbig-Haro flows, ][]{1997Reipurth_aspin}, but these have been mostly serendipitous discoveries while observing samples of YSOs \citep[e.g.][]{1998Sandell}.

Over the last 15 years,  multi-epoch, optical \citep[Gaia, ZTF, Pan-STARRS, ASSASN,][]{2016Chambers,2018Bellm,2018Jayasinghe,2021Hodgkin}, near-IR \citep[UKIDSS GPS, VVV, PGIR,][]{c2008Lucas,2012Saito,2016Moore}, mid-IR \citep[Spitzer, WISE, NEOWISE,][]{2003Benjamin,2010Wright,2014Mainzer} and sub-mm \citep[JCMT,][]{2021Lee,2024Mairs} surveys have led to an increase in the observations of all type of variables, including the rare detection of high-amplitude, long-lasting outbursts \citep{2018Hillenbrand,2021Guo,2024Guo_a,2024Lucas}. FUor and FUor-like YSOs, however, are still a rare class of variable stars, with roughly $\sim$50 objects classified as such over the last 85 years \citep{2018Connelley,2023Fischer}.  The lack of continuous, long-term monitoring in SFRs can partly explain the scarcity of FUor detection. In addition, until recently, monitoring has been done mostly at optical wavelengths, therefore looking at more evolved YSOs where the rate of FUor outbursts is low \citep{2019Contreras}. Observational and theoretical studies suggest that FUor outbursts are more frequent during the earlier Class 0 and Class I stages of young stellar evolution \citep[][]{2014Bae,2019Contreras,2019Hsieh, 2019Fischer,2021Park, 2022Zakri,2024Contreras}. YSOs at these stages are usually invisible at optical wavelengths and more readily observed at wavelengths longer than 1 $\mu$m.

Accurate statistics of FUor outbursts are critical to understanding the episodic accretion phenomena. In this paper, we aim to detect a population of FUors that may have gone undetected in the past, by analysing the multi-epoch near- to mid-IR photometric data of the objects in the Spitzer/IRAC Candidate YSO Catalog for the Inner Galactic Midplane \citep[SPICY,][]{2020Kuhn}. In Section \ref{sec:detfuors} we provide a description of the YSO sample, and the multi-epoch near- to mid IR surveys that will be used in our analysis. Section \ref{sec:lc_class} shows the method that is used to classify light curves with the goal of detecting long-term rises and decays that might be associated with FUors. In Section \ref{sec:candidate} we describe the characteristics of 717 candidate FUors that arise from the analysis of Section \ref{sec:lc_class}. In addition, in Section \ref{ssec:contamination} we present a discussion on possible sources of contamination. Section \ref{sec:follow-up} describes the follow-up observations of a sub-sample of candidate FUors. Section \ref{sec:results_spec} contains the discussion of the spectroscopic data and the confirmation of the majority of this sample as eruptive YSOs. Using the results of our analysis we provide, in Section \ref{sec:timescales}, some insights into the timescales involved in FUor outbursts. Finally Section \ref{sec:summary} summarises of our results.

\section{Detecting FUors}\label{sec:detfuors}

The number of FUor outbursts that we can detect depends both on the number of YSOs and on the amount of time that this sample is monitored \citep{2015Hillenbrand,2019Contreras}. In this work, we attempt to maximise these variables by 1) using the largest available catalogue of YSOs, the {\it Spitzer}/IRAC Candidate YSO (SPICY) Catalog for the Inner Galactic Midplane \citep{2020Kuhn}, and 2) using the multi-epoch (9--12 years) near- to mid-IR observations from the Vista Variables in the Via Lactea survey and its extension (VVV/VVVX), and the Wide-field Infrared Survey Explorer (WISE).

\subsection{The sample}\label{ssec:sample}

To develop the SPICY catalogue, \citet{2020Kuhn} used a random forest classification to select YSOs from {\it Spitzer} photometry obtained during the cryogenic mission. This includes seven {\it Spitzer}/IRAC surveys covering 613 square degrees. Their selection is also augmented with three near-infrared surveys: 2MASS \citep{2006Skrutskie}, UKIDSS Galactic Plane Survey \citep{2007Lawrence,c2008Lucas} and VVV \citep{2012Saito}. The SPICY catalog contains 117,446 YSO candidates.

The Spectral Energy Distribution (SED) of YSOs at wavelengths between 2 $< \lambda <$ 24 $\mu$m is generally used to determine the evolutionary stage of the system \citep[e.g.][]{1987Lada, 2014Dunham}. The different YSO Classes or stages are defined according to the value of their infrared spectral index, $\alpha$, defined as
\begin{equation}
\alpha = \frac{dlog(\lambda F_{\lambda})}{dlog\lambda},
\label{eq:sed}
\end{equation}
\noindent from which objects are classified as either Class 0, I, II or flat-spectrum YSOs \citep{1987Lada, 1993Andre, 1994Greene}.

Since the value of $\alpha$ estimated using eqn \ref{eq:sed} can be affected by reddening, \citet{2020Kuhn} uses 4.5$-$24 (from {\it Spitzer}) or $4.5-22 \mu$m (from {\it Spitzer} and {\it WISE}) colours to derive $\alpha$ in their sample. \citet{2020Kuhn} find 15943 (14\% of the sample) Class I ($\alpha >0.3$), 23810 (20\%) flat-spectrum ($-0.3<\alpha<0.3$), 59949 (51\%) Class II ($-1.6<\alpha<-0.3$), and 5352 (5\%) Class III ($\alpha<-1.6$) YSOs, with an additional 12392 (10\%) YSOs having uncertain classification. For the purposes of this work we select YSOs classified as either Class I, Class II or flat-spectrum sources, reducing the sample to 99,702 sources. 

\begin{figure*}
	\resizebox{2\columnwidth}{!}{\includegraphics[angle=0]{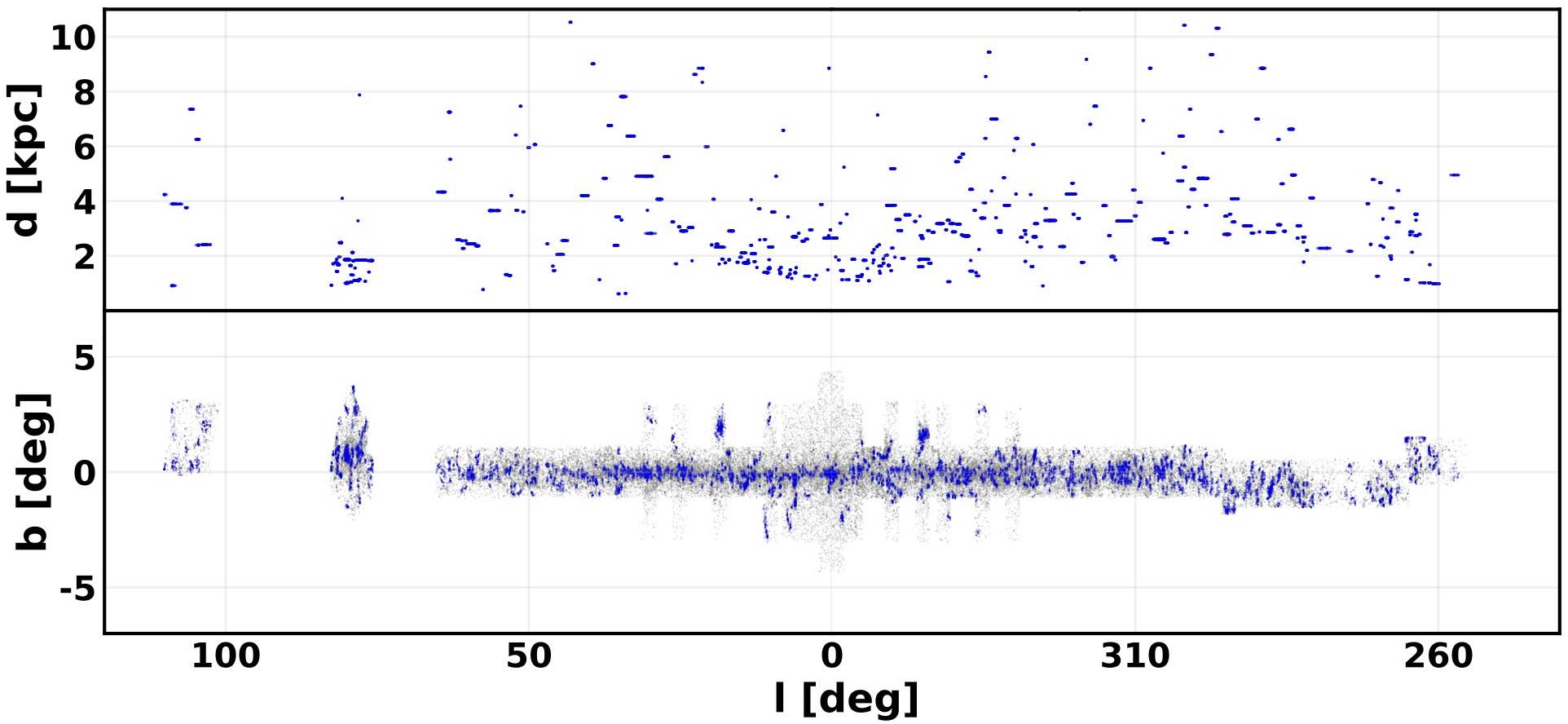}}
	 \caption{(top) Distance versus Galactic longitude for SPICY sources with avaialable distance information. (bottom) Galactic latitude versus longitude of SPICY sources (grey dots). Objects with available distance measurements are marked with blue dots.}
	 \label{fig:info_sample}
\end{figure*}

The overall sample of YSOs is distributed within the area covered by the {\it Spitzer} surveys, but with a higher density towards the Galactic midplane and some stellar associations (such as the Cygnus-X star-forming region at $l\sim78^{\circ}$, see Fig.\ \ref{fig:info_sample}). Information on typical distances for the YSOs in our sample can be determined from the mean parallax of stellar associations found by \citet{2020Kuhn}. 47,218 YSOs are associated with such stellar groups. Figure \ref{fig:info_sample} shows that the majority of the sample is located at distances larger than 0.6 kpc and  with a median value of 2.8 kpc.

\subsection{WISE}

This work uses mid-IR photometry from all-sky observations of the {\it WISE} telescope. {\it WISE} surveyed the entire sky in four bands, W1 (3.4 $\mu$m), W2 (4.6 $\mu$m), W3 (12 $\mu$m), and W4 (22 $\mu$m), with the spatial resolutions of 6.1\arcsec, 6.4\arcsec, 6.5\arcsec, and 12\arcsec, respectively, from 2010 January to September \citep{2010Wright}. The survey continued as the NEOWISE Post-Cryogenic Mission, using only the W1 and W2 bands, for an additional 4 months \citep{2011Mainzer}. In September 2013, WISE was reactivated as the NEOWISE-reactivation mission \citep[NEOWISE-R,][]{2014Mainzer}. NEOWISE-R stopped operating in 2024, and the latest released data set contains observations until 31 July 2024. For each visit to a particular area of the sky, {\it WISE} performs several photometric observations over a period of $\sim$few days. Each area of the sky is observed similarly every $\sim$ 6 months.

For the analysis of SPICY YSOs we used all the available data from the {\it WISE} telescope for observations from 2010 until 2021. The single-epoch data were collected from the NASA/IPAC Infrared Science Archive (IRSA) catalogues using a 3\arcsec\ radius from the coordinates of the YSO. For each source, we averaged the single epoch data taken over a few days to produce one epoch of photometry every six months \citep[following the procedures described by][]{2021Park}.

From the 99,702 SPICY sources in our sample (Section \ref{ssec:sample}), we find 53,662 objects (or 54\% of the sample) with WISE data (defined as having more than 3 epochs in W1 or W2 bands).

\subsection{VVV/VVVX}

For YSOs located in the area covered by the Vista Variables in the Via Lactea survey and its extension (VVV/VVVX) we also use a preliminary version of the VVV/VIRAC2 catalogue \citep{2018Smith,2025Smith}. The catalogue contains the Z, Y, J, H and K$_{\rm s}$ profile fitting photometry from the 2010 through 2019 observations for $\sim$500 million sources located in the original region covered by the VVV survey \citep{2010Minniti}. The astrometric properties of the VISTA K$_{\rm s}$ data also allow us to determine proper motions and parallaxes for VVV sources. The PSF photometry is derived using DoPHOT \citep{1993Schecter,2012Alonso} where a new absolute photometric calibration is obtained to mitigate issues that arise from blending of VVV sources in 2MASS data of crowded inner Galactic bulge regions \citep{2020Hajdu,2025Smith}.

The VIRAC2 catalogue also provides values for the proper motions of individual sources, which are calibrated using the Gaia absolute reference frame. Using the proper motions, probability density functions for stellar distance can be estimated using the method described in \citet{2021Guo} by adopting a Galactic rotation curve.

We have photometric information from the VIRAC2 catalogue for 32,092 objects (32\%). Combined, 16,546 sources (17\%) have WISE$+$VIRAC2 data. Inspection of VIRAC2 sources with no WISE counterparts reveals that they tend to have fainter K$_{\rm s}$ magnitudes than those sources with WISE$+$VIRAC2 data. It is likely that fainter sources fall below the detection limits of the {\it WISE} telescope.

\begin{figure*}
	\resizebox{2\columnwidth}{!}{\includegraphics[angle=0]{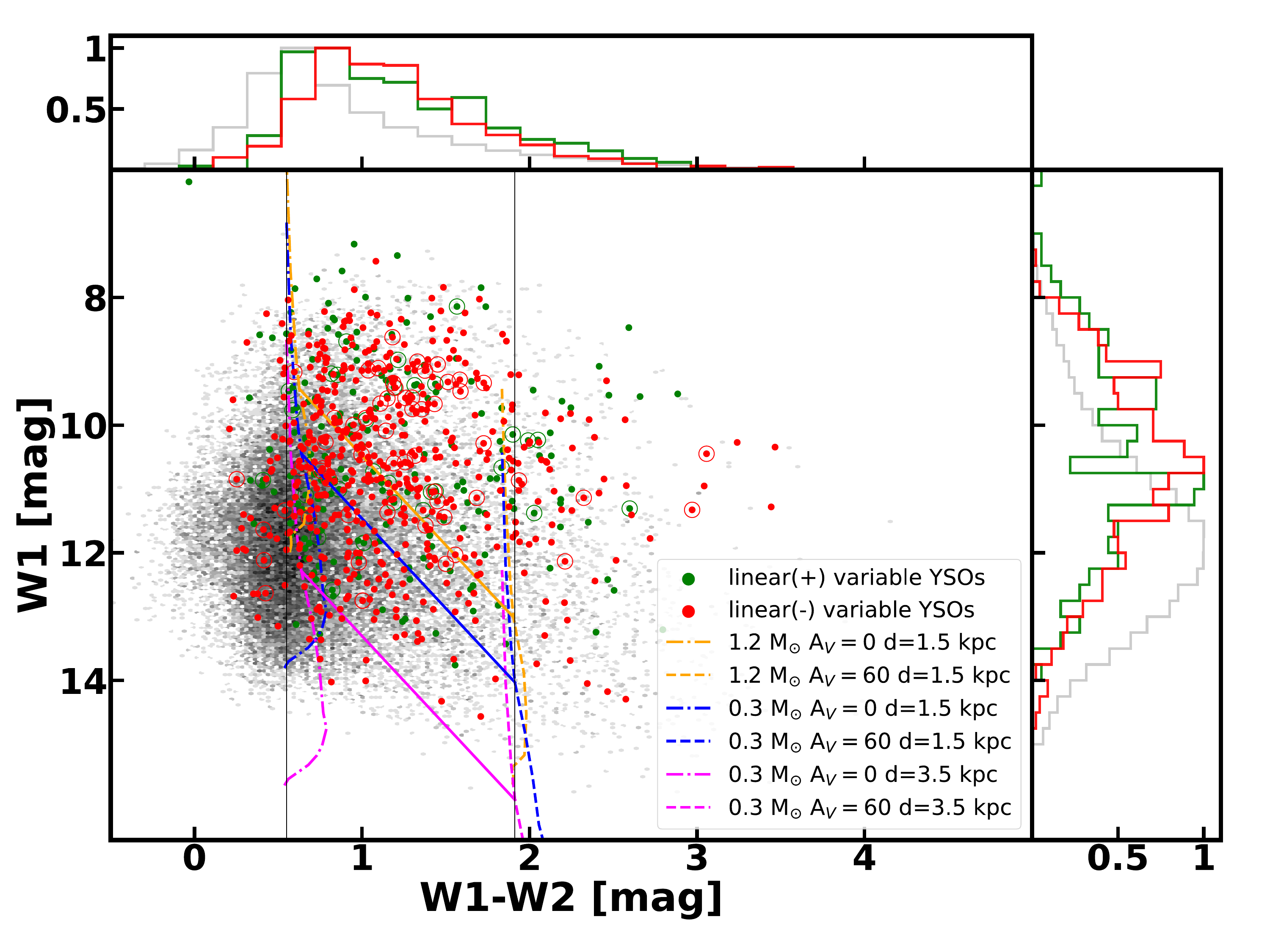}}
	 \caption{W1 vs W1$-$W2 CMD for SPICY YSOs (grey symbols). YSOs that are classified as {\it linear(-)} from their mid-IR light curves are shown as red filled circles, while {\it linear(+)} YSOs are shown in green filled circles. YSOs that have near-IR light curves with {\it linear(+)} or {\it linear(-)} classification are marked by red and green open circles, respectively. Isomass curves for an FUor outburst, as the mass accretion rate increases, are shown for a YSO with mass of the central star of M=0.3M$_{\odot}$ and M=1.2M$_{\odot}$. The curves, taken from \citet{2022Liu}, are artificially set at distances of 1.5 and 3.5 kpc and are shown for A$_{V}=$0 and A$_{V}=$60 mag. The blue, pink and orange solid lines connecting the curves, mark the point where the disk dominates emission. The vertical black solid lines roughly mark the colors of systems with A$_{V}=0$ and 60~mag from the \citet{2022Liu} models.}
	 \label{fig:cmd1}
\end{figure*}

\begin{figure*}
	\resizebox{2\columnwidth}{!}{\includegraphics[angle=0]{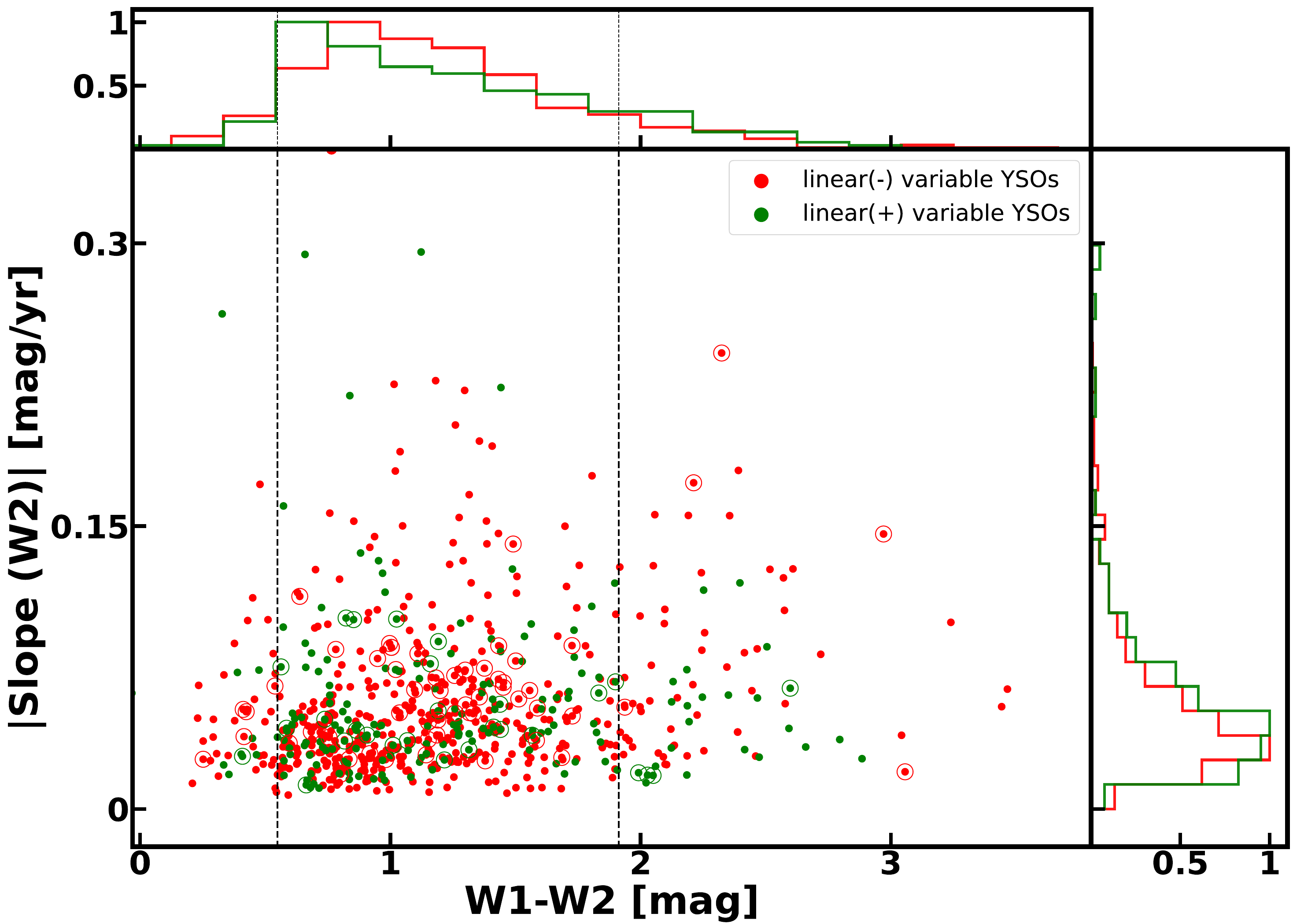}}
	 \caption{Average change per year (in magnitudes) vs. $W1-W2$ colour for YSOs that are classified as {\it linear}. The vertical black solid lines roughly mark the colors of systems with A$_{V}=0$ and 60~mag from the \citet{2022Liu} models. Symbols are the same as in Fig. \ref{fig:cmd1}.}
	 \label{fig:slope}
\end{figure*}

\section{Classification of light curves.}\label{sec:lc_class}

We identify candidate FUors with either slow-rising outbursts or showing evidence of continual dimming after reaching their peak brightness. For this purpose we use the methods developed by \citet{2021Park} to characterize the NEOWISE lightcurves. 

\citet{2021Park} search for secular and stochastic variability in the long-term data of $\sim$7000 YSOs. Candidate variable stars are selected if they show $\Delta W2/\sigma_{W2}\geq3$\footnote{$\Delta$ is defined as the maximum-minimum magnitude, and $\sigma$ is calculated by adding, in quadrature, the mean error and the standard deviation (in magnitudes) of the exposures in each epoch}. Lomb-Scargle periodogram \citep[LSP, ][]{1976Lomb,1989Scargle} and linear fits are used to search for secular trends in the light curves of candidate variable stars. To quantify the uncertainty in a particular LSP peak or best-fit linear slope, \citet{2021Park} define a false alarm probability (FAP) following the method developed by \citet{2008Baluev} and the methods used in the analysis of sub-mm light curves by \citet{2021Lee}. YSOs that are well fitted by a linear regression model (or FAP$_{lin}<10^{-4}$) are classified as {\it linear($+$)}, if the source brightens over time, or {\it linear($-$)} if the source becomes fainter with time. The remaining YSOs are further classified into secular (periodic, curved) or stochastic (burst, drop, irregular) variables according to a selection of criteria defined by \citet{2021Park}. 

The aim to identify long-term FUor outbursts already discards the light curves falling into a classification as stochastic or periodic (P$<1400$~d). Inspection of figure 7 in \citet{2021Park} shows that the {\it linear(-)} and {\it linear(+)} categories identified in that work are an ideal place to search for the slow-rising outbursts and those that show continuous dimming after reaching peak brightness. YSOs classified as curved could contain some outbursts, but likely of intermediate duration. Previous experience in the analysis of mid-IR light curves show that YSOs that are well fitted by a sinusoidal curve with long periods (P$\geq$3000~d) already fall in the linear category defined in \cite{2021Park}. 

We searched for objects that fall in the {\it linear} classification of \citet{2021Park}, by analysing the mid-IR light curves of the 99702 SPICY objects. Firstly, we selected YSOs with at least 12 epochs in both W1 and W2. The latter condition is fulfilled by 45636 YSOs ( or 46\% of the sample). From these, 37847 YSOs, or 38 \% of the sample, fulfill either $\Delta_{W1}/\sigma_{W1}\geq3$ or $\Delta_{W2}/\sigma_{W2}\geq3$. Finally, we find 1087 and 2043 {\it linear} YSOs in W1 and W2, respectively.

We also explored the available data at near-IR (K$_{s}$) wavelengths. However, we are unable to use exactly the same methods as \citet{2021Park} to classify the light curves at these wavelengths. Using the same threshold for the FAPs as \citet{2021Park} yields almost every variable candidate being classified as {\it linear}. The higher cadence of VVV observations may lead to this problem, as discussed below.

The FAPs are defined following \citet{2008Baluev}, as
\begin{equation}
    FAP_{A}=(1-P_{A})^{((N_{epochs}-D_{A})/2)}
\end{equation}
\noindent with $D_{A}$ the number of parameters in the model and P$_{A}$ the statistical power of the best-fit ``A'' hypothesis, defined as:

\begin{equation}
    P_{A}=\frac{\chi^{2}_{N}-\chi^{2}_{A}}{\chi^{2}_{N}}
\end{equation}

\noindent with $\chi^{2}_{N}$ and $\chi^{2}_{A}$ the chi-squared values of the non-varying and ``A'' hypothesis, respectively. The values of the FAPs decrease with increasing number in degrees of freedom (or increasing epochs of observations). Due to this dependence and the larger number of epochs in VVV observations, FAP$_{lin}$ and FAP$_{LSP}$ become too low and reach almost zero, leading to almost every variable candidate classified as {\it linear} in the near-IR.

Objects that are classified as {\it linear} or {\it periodic} from their mid-IR WISE data, show values of the statistical power $P_{A}$, defined above, that are above $P_{A}\sim0.7$. 

For the VIRAC2 near-IR light curves, we calculate $P_{A}$ for the best-fit linear ($P_{A, lin}$) and periodic ($P_{A, LSP}$) models. From visual inspection of light curves with $P_{A}>0.7$ we determine that objects with values above this threshold are well fitted by linear or periodic models. Then, the classification for optical and near-IR light curves that show $\Delta/\sigma\geq3$ follows the criteria:

\begin{align*}
    \mathrm{linear}\quad &P_{A,lin}\geq0.7\\
    \mathrm{periodic}\quad &P_{A,lin}<0.7\quad P_{A,LSP}\geq0.7\quad P<1400~\text{d}\\
    \mathrm{curved}\quad &P_{A,lin}<0.7\quad P_{A,LSP}\geq0.7\quad P\geq1400~\text{d}
\end{align*}

\noindent with the remaining YSOs with $\Delta/\sigma\geq3$ being classified into the different stochastic classes following the same criteria as \citet{2021Park}.

From the group of candidate variable YSOs, 717 (0.72\% of the sample) are well-fitted by a linear model in both W1 and W2, and are therefore classified as {\it linear}. From these, 526 have negative slopes ({\it linear(-)}), while 191 show positive slopes ({\it linear(+)}). From the 717 {\it linear} objects identified from mid-IR data, 235 have near-IR data from the VVV survey. We find that 86 of them are also classified as {\it linear} from the 2010-2019 K$_{\rm s}$ data.  From 191 {\it linear(+)} YSOs, 28 have the same classification from their VVV light curves. In the case of 526 {\it linear(-)} objects, 57 are classified as {\it linear(-)} in K$_{\rm s}$. Only in one object we find classification as {\it linear(-)} in the mid-IR while having a {\it linear(+)} in the near-IR (SPICY 43479). Although the different classification may arise from non-contemporaneous coverage between WISE/NEOWISE and VVV, we cannot discard that the anti-correlation could be associated with structural changes of the inner accretion disk \citep{2019Bryan}.

The variability parameters and classification for the mid-IR light curves of 717 {\it linear} YSOs are presented in Table \ref{tab:allvar}. The same parameters arising from near-IR light curves are also presented, when available.

\movetableright=-2in
\begin{table*}
\caption{Information on the 717 SPICY YSOs classified as {\it linear} from their mid-IR light curves. The displayed version only contains the 20 YSOs with spectroscopic data that are discussed in the main text. The full table is available online.\label{tab:allvar}}
\resizebox{\textwidth}{!}{
\begin{tabular}{lccccccccccccccccccccc} 
\tableline
SPICY ID & ra (J2000) & dec (J2000) & YSO class &  I1 & I1err & I2 & I2err & W1 & W1err & W2 & W2err & K$_{s}$ & K$_{s}$err & $\Delta$W1 & $\Delta$W2 & $\Delta$K$_{s}$ & N$_{W1}$ & N$_{W2}$ & N$_{Ks}$ &  Var class ($W1$,$W2$) & Var class (K$_{s}$) \\
\tableline
11492 & 11:43:09.5 & -62:21:13.3 & ClassII & 13.40 &  0.07 & 12.42 &  0.11 & 11.78 &  0.82 & 10.75 &  0.76 & 15.08 &  0.06 &  2.37 &  2.30 &  4.32 &   16 &   16 & 126 & linear(+) & irregular \\
 15180 & 12:57:44.2 & -62:15:06.4 & FS & 12.27 &  0.04 & 10.98 &  0.05 & 10.37 &  0.60 &  9.23 &  0.46 & 12.98 &  0.02 &  2.23 &  1.70 &  2.47 &   18 &   18 & 158 & linear(+) &linear(+) \\
 15470 & 13:01:20.7 & -62:20:01.6 & ClassII &  9.55 &  0.04 &  8.65 &  0.04 &  9.89 &  0.15 &  8.72 &  0.12 & 11.86 &  0.02 &  0.52 &  0.46 &  0.60 &   18 &   18 & 149 & linear(-) &linear(-) \\
 21349 & 14:12:48.7 & -61:22:50.6 & ClassII &  8.97 &  0.03 &  8.24 &  0.04 &  9.48 &  0.24 &  8.40 &  0.25 & 11.12 &  0.02 &  0.80 &  0.85 &  0.48 &   17 &   17 & 111 & linear(-) &irregular \\ 
 29017 & 15:45:18.4 & -54:10:36.9 & ClassII & 12.16 &  0.07 & 11.40 &  0.07 &  9.57 &  0.43 &  8.70 &  0.41 & 11.70 &  0.02 &  1.97 &  1.88 &  2.83 &   17 &   17 & 133 & linear(+) &linear(-) \\ 
 31759 & 15:59:26.3 & -51:57:11.5 & ClassII & 13.36 &  0.09 & 12.35 &  0.11 & 10.31 &  0.29 &  9.16 &  0.33 & 12.52 &  0.03 &  0.90 &  0.86 &  4.33 &   14 &   14 & 360 & linear(-) &irregular \\ 
 35235 & 16:18:24.8 & -48:54:32.1 & ClassII &  9.91 &  0.06 &  9.23 &  0.05 & 10.36 &  0.21 &  9.36 &  0.21 & 11.66 &  0.02 &  0.71 &  0.68 &  0.46 &   17 &   17 & 124 & linear(-) &linear(-) \\ 
 36590 & 16:23:27.1 & -49:44:43.6 & FS & 12.56 &  0.07 & 11.36 &  0.07 & 11.02 &  0.31 &  9.76 &  0.31 & 13.67 &  0.03 &  1.08 &  1.12 &  3.80 &   14 &   14 & 123 & linear(-) &curved \\ 
 42901 & 16:51:57.8 & -45:42:39.3 & FS &  9.27 &  0.04 &  8.29 &  0.04 &  9.18 &  0.11 &  8.18 &  0.19 & 11.58 &  0.02 &  0.42 &  0.64 &  0.93 &   17 &   17 & 191 & linear(-) &irregular \\ 
 57130 & 17:34:23.1 & -30:52:23.3 & ClassII &  8.76 &  0.05 &  7.67 &  0.04 &  9.26 &  0.17 &  7.85 &  0.15 & 11.81 &  0.02 &  0.54 &  0.45 &  0.68 &   17 &   17 & 189 & linear(-) &linear(-) \\ 
 63130 & 17:46:33.8 & -29:22:44.9 & ClassI & 10.79 &  0.05 &  8.95 &  0.04 &  9.86 &  0.59 &  7.87 &  0.28 & 14.93 &  0.06 &  2.07 &  0.94 &  4.54 &   17 &   17 & 225 & linear(+) &irregular \\
 65417 & 17:48:26.3 & -24:07:33.2 & ClassII &  8.93 &  0.04 &  8.15 &  0.03 &  9.66 &  0.18 &  8.55 &  0.19 & 11.20 &  0.03 &  0.74 &  0.75 &  0.83 &   17 &   17 & 289 & linear(-) &linear(-) \\ 
 68600 & 17:55:01.0 & -28:01:27.4 & ClassII &  9.31 &  0.04 &  8.43 &  0.04 &  9.49 &  0.16 &  8.38 &  0.15 & 11.32 &  0.03 &  0.57 &  0.48 &  0.96 &   17 &   17 & 196 & linear(-) &linear(-) \\ 
 68696 & 17:55:15.3 & -28:52:58.3 & ClassII &  9.01 &  0.02 &  8.22 &  0.03 &  9.64 &  0.13 &  8.46 &  0.11 & 11.74 &  0.02 &  0.48 &  0.46 &  0.62 &   17 &   17 & 191 & linear(-) &linear(-)\\ 
 87984 & 18:36:46.3 & -01:10:29.5 & ClassI &  8.73 &  0.03 &  8.01 &  0.05 &  9.17 &  0.18 &  8.00 &  0.20 & -- & -- &  0.51 &  0.60 & -- &   17 &   17 & -- &  linear(-) & -- \\ 
 95397 & 18:57:20.3 & +01:57:12.2 & ClassII & 11.48 &  0.04 & 10.55 &  0.05 &  7.68 &  0.79 &  6.80 &  0.69 & -- & -- &  3.55 &  3.24 & -- &   16 &   16 & -- &  linear(+) &-- \\ 
 99341 & 19:11:38.8 & +09:02:59.1 & ClassII & 12.67 &  0.06 & 11.73 &  0.07 &  9.38 &  0.70 &  8.43 &  0.65 & -- & -- &  3.03 &  2.84 & -- &   16 &   16 & -- &  linear(+) &-- \\ 
100587 & 19:17:17.9 & +11:16:32.3 & ClassII & 11.12 &  0.05 & 10.44 &  0.06 & 11.44 &  0.43 & 10.52 &  0.46 & -- & -- &  1.34 &  1.50 & -- &   16 &   16 & -- &  linear(-) &-- \\ 
109102 & 20:23:57.1 & +38:51:39.7 & FS & 11.09 &  0.03 & 10.37 &  0.03 & 11.30 &  0.02 & 10.37 &  0.03 & -- & -- &  0.11 &  0.13 & -- &   17 &   17 & -- &  linear(-) &-- \\ 
111302 & 20:30:55.7 & +38:40:23.1 & ClassII &  9.14 &  0.03 &  8.55 &  0.04 &  8.96 &  0.15 &  8.21 &  0.18 & -- & -- &  0.48 &  0.56 & -- &   17 &   17 & -- & linear(-) & -- \\ 
\tableline
\end{tabular}
    }
    
\end{table*}


\subsection{Candidate FUors}\label{sec:candidate}

In Fig.~\ref{fig:cmd1}, we show the W1 versus W1$-$W2 color-magnitude (CMD) diagram for SPICY YSOs with available WISE photometry, as well as the location of 717 YSOs classified as {\it linear}. In the same figure we also show the location of FUor outbursts in the CMD, as predicted by the SED models of FUor outbursts from \citet{2022Liu}. We compare with models for a mass of the central star of M$_{\star}=0.3$ M$_{\odot}$ and M$_{\star}=1.2$ M$_{\odot}$. The magnitudes and colors are corrected for distances of 1.5 and 3.5 kpc, and extinction values in the V-band of A$_{V}=0$ and A$_{V}=60$~mag. The extinction at the wavelengths of the WISE filters are estimated using the \citet{2021Gordon} Milky Way extinction curves. In the figure, we also mark the point in the iso-mass tracks of FUor outbursts at which the disk dominates the emission from the YSO. We find that FUors can be located over a wide range of magnitudes, depending on the distance to the source. The colors of these systems, however, are mostly contained in the region $0.55<W1-W2<1.91$~mag.

Figure~\ref{fig:cmd1} shows that the distribution of $W1-W2$ colour of {\it linear} sources appears to peak at redder values compared with the distribution for the overall sample of SPICY objects. We find that 92\% of the {\it linear} YSOs are located at $W1-W2 > 0.55$~mag \citep[the lower limit defined by the ][models]{2022Liu}. If we take $W1-W2\simeq1.91$~mag, the colour defined by \citet{2022Liu} models with $A_{V}=60$~mag \citep[close to the maximum value of A$_{V}$ for known FUors in][]{2018Connelley}, as an upper limit to define the region where FUors are preferentially located, we find that 82\% of {\it linear} objects are in this region. The {\it linear} sources are also located at brighter magnitudes compared with the overall sample of protostars, which would agree with the expectations of these objects being candidate FUors. However, we cannot discard the possibility that this is associated with our inability to classify fainter sources with larger uncertainties in magnitude. Similar distributions and conclusions are derived when only considering the objects that have a {\it linear} classification from VVV data.

Differences also exist among the YSOs classified as {\it linear}. In Fig.~\ref{fig:slope} we show the absolute value of the average change in magnitude per year, versus the $W1-W2$ colour for the {\it linear} YSOs. Here we find that the majority of the sources show shallow slopes, with changes in most sources ($\sim90\%$) located at values lower than 0.1 mag yr$^{-1}$. The visual inspection of Fig.~\ref{fig:slope} seems to indicate that  {\it linear(+)} YSOs have a distribution that is slightly skewed towards bluer $W1-W2$ colours and a slightly higher fraction of objects with larger values of the slope. We evaluate these statements by performing a set of different statistical tests (Kolmogorov-Smirnov, Wilcoxon and Anderson-Darling) on these distributions. The tests support the differences in colour distribution between the two samples (although with some dependence on the chosen bin size). However, we cannot conclude that the two samples show differences in the distribution of the slope.

The validity of any analysis based on the sample defined above depends on whether all objects in our {\it linear} sample are FUor outbursts. Confirming this requires analysis of the expected degree of contamination 
as well as spectroscopic follow-up of our sample.

\subsection{Contamination}\label{ssec:contamination}

The photometric characteristics of the {\it linear} sources, discussed in the previous section, are in line with the expectations of FUors and can therefore be classified as candidate FUors. However, we also need to consider the possibility of other classes of variable stars contaminating our sample \citep[see e.g.][]{2021jelee}.

\citet{2020Kuhn} discuss the possibility of evolved stars and extragalactic sources as the two main sources of contamination in their sample. Both types of objects can mimic the near- to mid-IR colours of YSOs and therefore contaminate YSO searches \citep[see e.g.][]{2008Robitaille}. The evolved sources can also complicate the interpretation of the long-term declines in our sample. Some symbiotic stars show similar long-term declines in their optical light curves \citep[see e.g. Figure 1 in][]{2009Gromadzki}.

\citet{2020Kuhn} expect a low level of contamination, as evolved and extragalactic sources tend to be located in areas that are flagged as contaminants in their analysis, and therefore are less likely to be included in the SPICY catalogue. \citet{2022Guo} finds 253 SPICY sources that show periodic light curves that are likely associated with pulsation in evolved stars. Meanwhile, \citet{2023Kuhn} spectroscopically assessed the contamination to be $<$10\% for the subset of optically bright SPICY sources. However, there is no formal assessment of contamination by \citet{2020Kuhn}, as this would require a more complete spectroscopic follow-up. 

FUors are identified due to their unique spectroscopic characteristics during outburst. These include the gradual change of spectral type with wavelength, a lack of emission lines, absorption from water vapor and other molecular bands, including the appearance of strong and broad $^{12}$CO $\nu=2-0$ absorption at 2.2935 $\mu$m \citep{2010Connelley, 2018Connelley}.  During the outburst, the accretion disk dominates emission in the system, where absorption lines arise due to a cooler disk surface compared with the viscously-heated midplane \citep[][]{1977Herbig,1996Hartmann,2022Liu}.

The near-IR spectroscopic characteristics of FUors are very similar to those of K-M giant stars. In addition the FUor spectra can, broadly speaking, resemble the mid- to late M-type atmospheres of young brown dwarfs \citep[e.g.][]{2018Connelley}. For the latter, we argue that given the typical distances of the SPICY sample, and the low luminosities of brown dwarfs, these objects would be located towards the faint magnitude limits of the WISE observations. Therefore it is unlikely that objects in our candidate FUors sample are brown dwarfs.

In the case of evolved stars, we can use additional evidence to discard contamination from these sources. For example, the observation of $^{13}$CO absorption is a useful indicator of the evolved nature of a source \citep{2017Contreras_a,2021Guo}. In addition, field giant stars show low rotation velocities, which would not lead to broadened absorption features. Only 2\% of giant stars investigated by \citet{2011Carlberg} show a projected rotational velocity, $v\sin i$, larger than 10 km s$^{-1}$; this fraction reduces to 0.7\% of the sample for $v\sin i>20$ km s$^{-1}$. However, giant stars in symbiotic systems show faster rotation than field giant stars, with some evolved objects in D-type symbiotic systems showing rotation with $v\sin i\sim30-40$ km s$^{-1}$. The infrared colours of these systems, especially D-type symbiotics,  overlap with those of YSOs in CMDs used for the classification of 
objects \citep[see e.g.][]{2008Corradi}.

High-resolution spectroscopy is useful to determine velocities from rotational or disk broadening \citep[e.g.][]{2020Park} and discriminate FUors from background giant stars. Using velocities to discriminate becomes more complicated for low- and mid-resolution spectroscopy.

The issue of potential contamination will be taken into account when analyzing the spectra of 20 candidates in the following sections.
 
\section{Follow-up of candidate FUors}\label{sec:follow-up}

From the list of candidate FUors we carried out spectroscopic follow-up of 10 sources that were bright enough to obtain moderate-to-high SNR observations with the Gemini/IGRINS and IRTF/Spex instruments. We were also able to find spectroscopic information in the literature for 10 additional sources. Information on the distance and evolutionary stage for the 20 sources is presented in Table \ref{tab:app} and Appendix \ref{sec:app1}.

\begin{figure}
	\resizebox{\columnwidth}{!}{\includegraphics[angle=0]{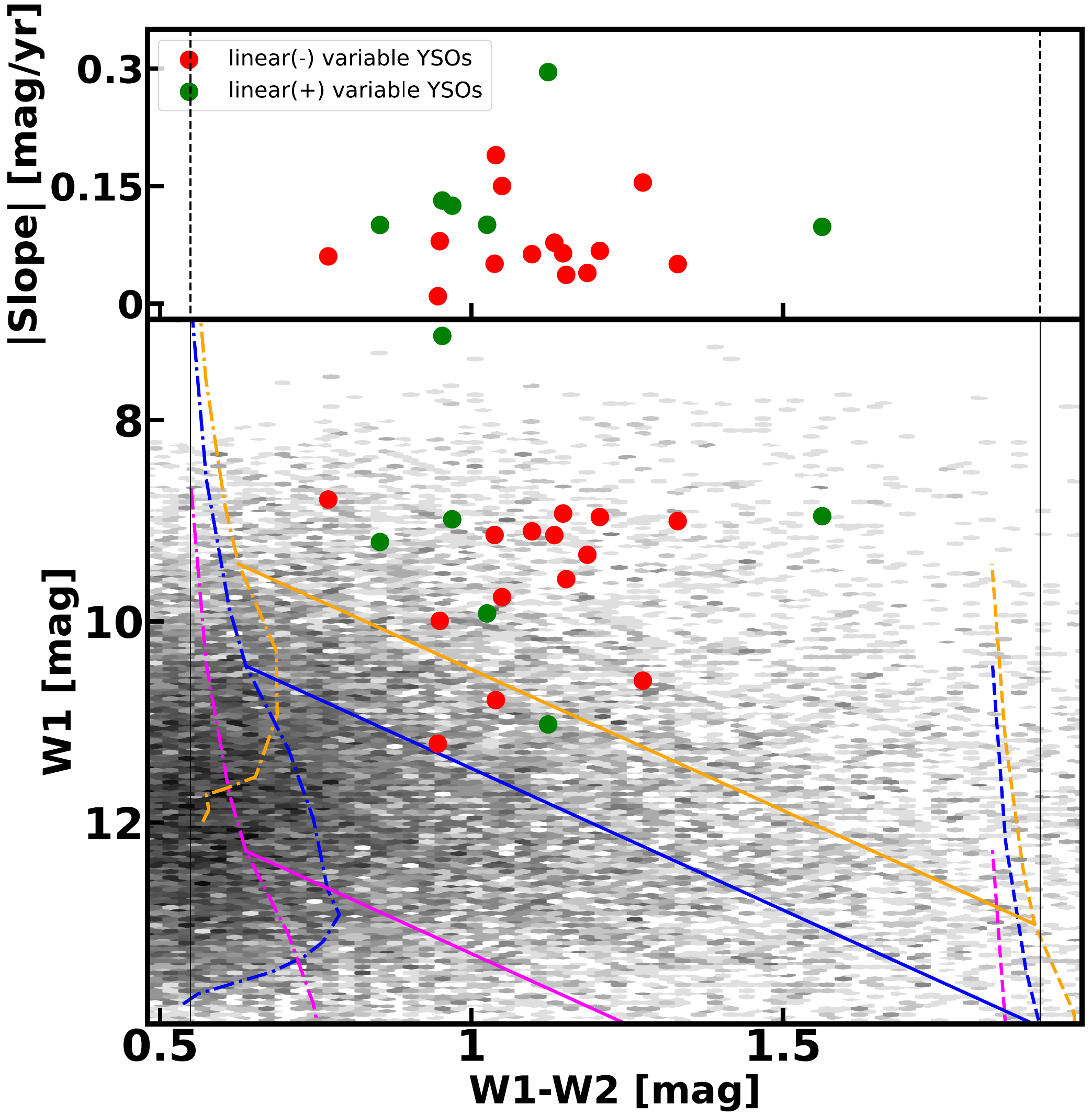}}
	 \caption{Location in average change per year (in magnitudes) vs. $W1-W2$ colour (top), and  $W1$ vs. $W1-W2$ colour-magnitude (bottom) diagram of 20 {\it linear} YSOs with spectroscopic data. Grey symbols and overlaid lines are the same as in Fig. \ref{fig:cmd1}.}
	 \label{fig:photprop}
\end{figure}

The locations of the 20 sources in the CMD and the slope-versus-colour diagram are shown in Fig.~\ref{fig:photprop}. We observe that the majority of sources are located within the regions defined by the \citet{2022Liu} theoretical models of FUor outbursts using extinction values of A$_{V}=0$ and 60~mag (marked by the black lines in the Figure). Most sources show shallow values of the slope in Fig.~\ref{fig:photprop}, with {\it linear(-)} objects tending to locate at lower values than objects classified as {\it linear(+)}.

\subsection{Spectroscopy from the literature}
 
In 10 objects we find near-IR spectroscopic data in the literature. These include IRTF/SpeX observations of SPICY 87984 \citep{2010Connelley} 95397, 99341 and 100587 \citep{2023Contreras_b}. In addition we find ESO VLT/X-shooter  observations for SPICY 11492, 31759 and 63130 \citep{2024Guo_a}. Finally, Magellan/FIRE observations of SPICY 15180, 29017 and 36590 were published by \citet{2017Contreras} and \citet{2021Guo}.

\subsection{IRTF/SpeX Spectroscopy}

We obtained near-IR spectra of SPICY 109102 and 111302 on 7 and 8 October 2023 (HST) using SpeX \citep{rayner03} mounted at the NASA Infrared Telescope Facility (IRTF) on Mauna Kea (programme 2023B079, PI Contreras Pe\~{n}a). The cross-dispersed spectra cover 0.7--2.5 $\mu$m  spectra at $R\sim 2000$, obtained with the 0.5\arcsec\ slit. Total integration times ranged from 480 to 840 s, with individual exposures of 60 s.  Bright A0V standard stars were observed for telluric calibration. All spectra were reduced and calibrated using Spextool version 4.1 \citep{cushing04}. 

\subsection{Gemini South/IGRINS Spectroscopy}

We observed 8 YSOs that were selected as candidate FUors with the Immersion Grating Infrared Spectrograph (IGRINS) installed at the 8.1 m Gemini South telescope between 2023 February 18 and 2023 March 17 (programme GS-2023A-Q-207. PI Lee, H. -G.). IGRINS simultaneously covers the H (1.49--1.80 $\mu$m) and K (1.96--2.46 $\mu$m) bands, with R$\simeq$45,000 \citep{2018Mace}. 

The observations were carried out on an ABBAABBA pattern, with individual exposure times between 310 and 360 s. The spectra have been reduced by the IGRINS pipeline\footnote{https://github.com/igrins/plp}. Standard A0 stars were observed immediately after the observation of each FUor candidate for purposes of telluric correction. The dates of observations and the standard star used for correction are presented in Table \ref{tab:spec}.

\movetableright=-0.1in
\begin{table}
	\centering
	\caption{Information on the eight SPICY YSOs with IGRINS observations.}
	\label{tab:spec}
\resizebox{\columnwidth}{!}{
   	\begin{tabular}{lccc} 
		\hline
 SPICY ID & Standard & Obs Date & Exp time (s) \\
\hline
 SPICY 15470 & HIP58898 & 19-Feb-2023 & 8$\times$360\\
  SPICY 21349 & HIP83066 & 15-March-2023 & 8$\times$310\\
   SPICY 35235 & HIP79477 & 15-March-2023 & 8$\times$360\\
  SPICY 42901 & HIP84267 & 18-March-2023 & 8$\times$340\\
   SPICY 57130 & HIP86098 & 16-March-2023 & 8$\times$360\\
  SPICY 65417 & HIP79332 & 19-Feb-2023 & 8$\times$320\\
   SPICY 68600 & HIP85871 & 17-March-2023 & 8$\times$340\\
  SPICY 68696 & HIP83399 & 17-March-2023 & 8$\times$360\\
\hline
	\end{tabular}}
\end{table}

\subsection{Historical light curves}\label{sssec:light_curves}

For the 20 sources with spectroscopic data we collected additional photometry from optical, near- and mid-IR surveys available through the VizieR catalogue access tool \citep{2000Oschsenbein}. For most cases, the oldest available data arises from the 2MASS or DENIS \citep{1994Epchtein} surveys. 

\begin{figure*}
	\resizebox{\textwidth}{!}{\includegraphics[angle=0]{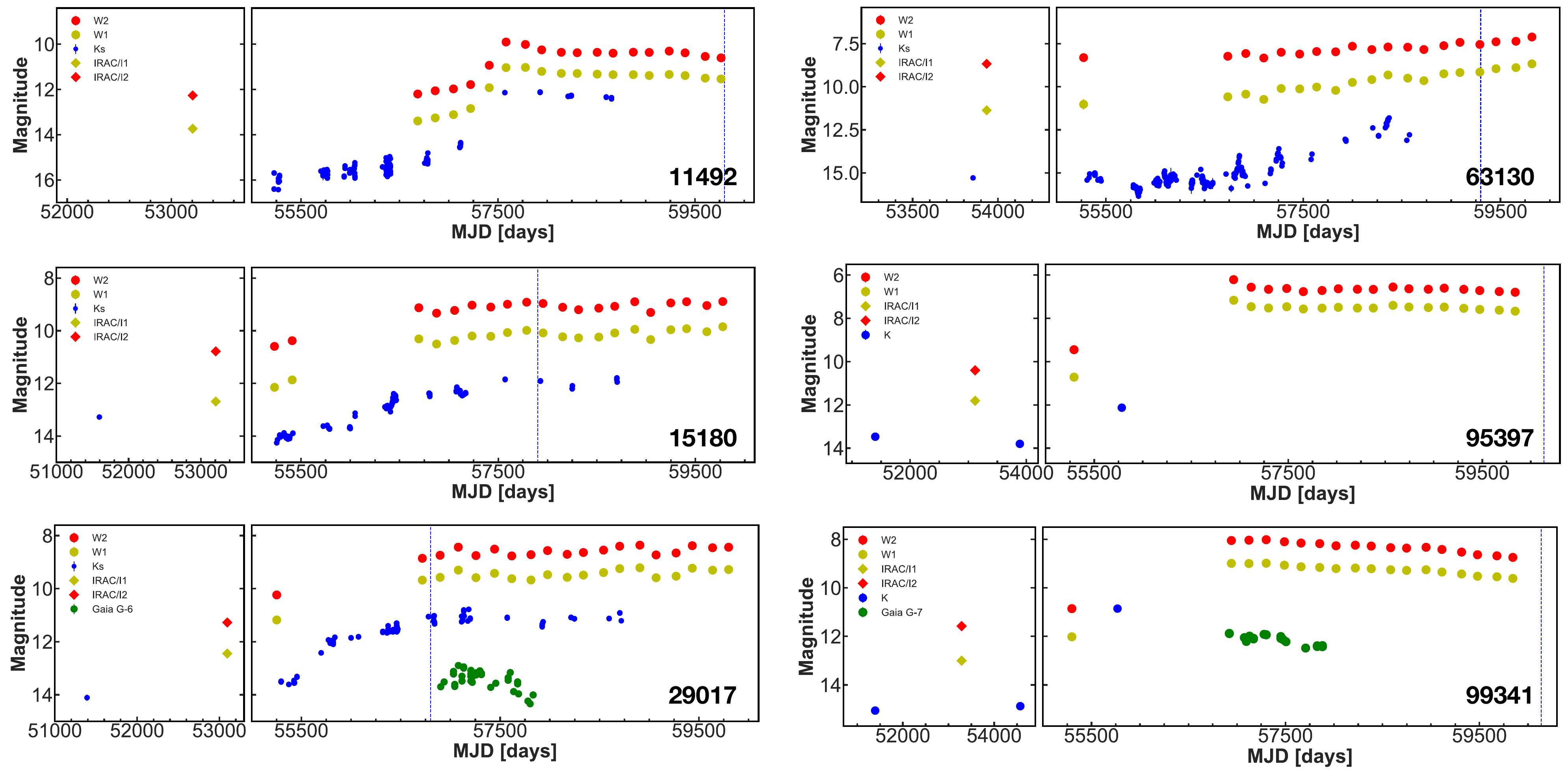}}
	 \caption{{\it Gaia} G (green), Near-IR K/K$_{\mathrm{s}}$ (blue), mid-IR W1/IRAC1 (yellow) and W2/IRAC2 (red) light curves of 6 {\it linear(+)} YSOs with spectroscopic data. The blue vertical line marks the date of the spectroscopic follow-up.}
	 \label{fig:yso2}
\end{figure*}

\begin{figure*}
	\resizebox{\textwidth}{!}{\includegraphics[angle=0]{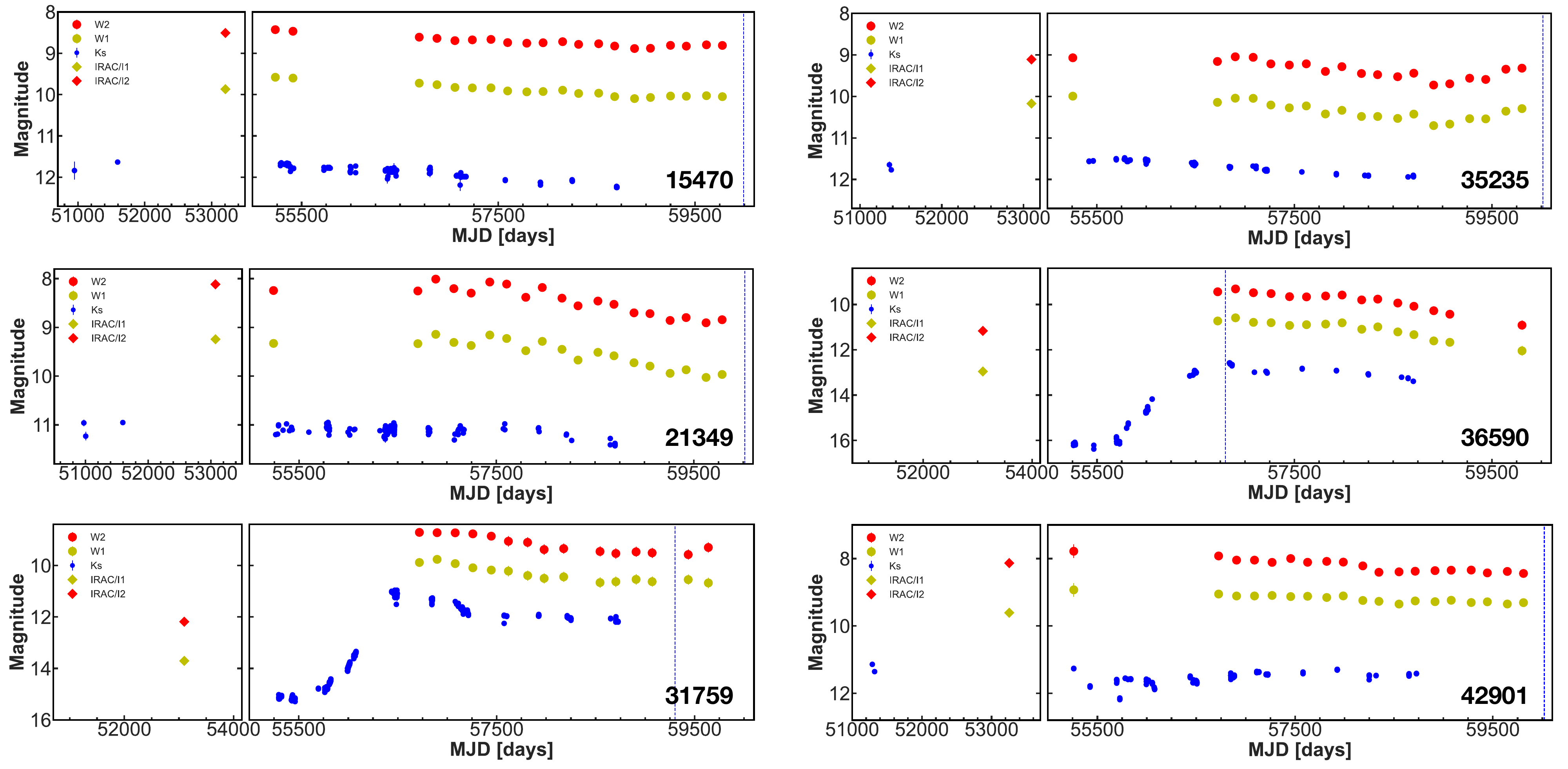}}\\
         \resizebox{\textwidth}{!}{\includegraphics[angle=0]{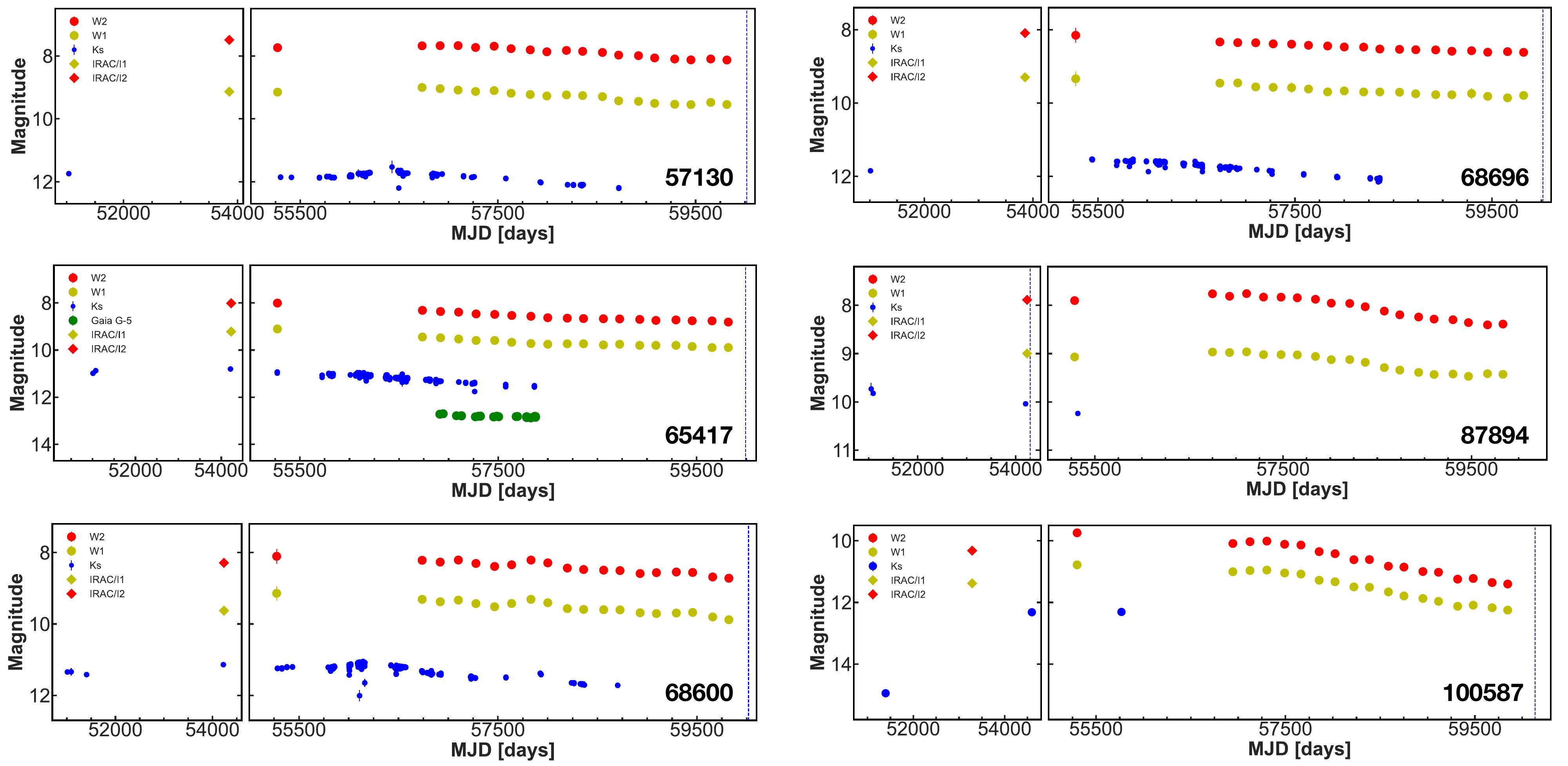}}\\
         \resizebox{\textwidth}{!}{\includegraphics[angle=0]{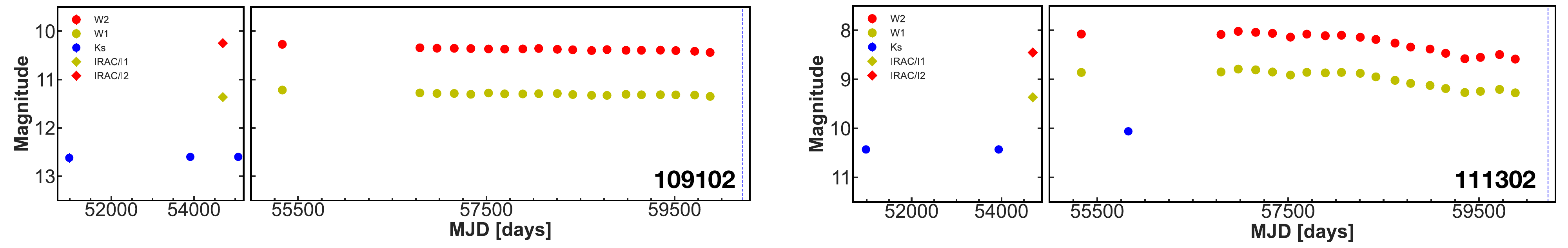}}
	 \caption{ Near-IR K/K$_{\mathrm{s}}$ (blue), mid-IR W1/IRAC1 (yellow) and W2/IRAC2 (red) light curves of 14 {\it linear(-)} YSOs with spectroscopic data. The blue vertical line marks the date of the spectroscopic follow-up.}
	 \label{fig:yso1}
\end{figure*}

In three objects, SPICY 29017 ({\it linear(+)}), 65417 ({\it linear(-)}), and 99341 ({\it linear(+)}) we have optical photometry from {\it Gaia}. For these objects we searched digitized photographic plate images from the SuperCOSMOS survey \citep{2001Hambly}. In SPICY 65417 the oldest photographic plate images were taken in 1985. Comparison with PanSTARRS r-band cutout image shows that the star was at a similar brightness level. This implies that, if this is truly a FUor, the object went into an outburst more than 38 years ago. In SPICY 99341 the photographic plate images were taken between 1950 and 1995, and in none of them we can see the source.  PanSTARRS r- and i-band cutout images confirm the outburst of the source around 2010-2011. A similar date for the outburst is reached when comparing WISE and UKIDSS data \citep{2023Contreras_b}. Finally, SPICY 29017  does not show up in the photographic plate images (1977-1991 observations). There is a faint source at the location of SPICY 29017 that shows up in r- and i-band images from the SkyMapper Southern Survey \citep{2019Onken} taken at similar dates to the {\it Gaia} observations, which would point to a recent outburst of the source. The source is brighter in z-band images from Skymapper, but the images taken in 2016 and 2017, show a decline in brightness which agrees with the {\it Gaia} light curve.  The light curves of these 6 {\it linear(+)} and 14 {\it linear(-)} objects with spectroscopic data are shown in Figs. \ref{fig:yso2} and \ref{fig:yso1}, respectively. 

\section{Spectroscopic confirmation of {\it linear} FUors}\label{sec:results_spec}

\subsection{Objects in the literature}

\textit{\textbf{Linear($-$)}}: SPICY 87984 (IRAS 18341$-$0113N) is part of the Class I sample of \citet{2010Connelley}. In their work, the near-IR spectrum of the source shows a triangular H-band continuum, $^{12}$CO absorption that is in excess for the best-fit spectral type for the YSO (K0-M4), and lack of emission lines. Given some of the similarities to FUor spectra, \citet{2010Connelley} note that this object might be experiencing weak FUor-like activity. 

SPICY 31759 \citep[L222\_33,][]{2024Guo_a}, SPICY 36590 \citep[VVVv270,][]{2017Contreras, 2017Contreras_a} and SPICY 100587 \citep{2023Contreras_b} all show high-amplitude outbursts ($\Delta~K_{\rm s}>2.5$~mag). The outbursts occurred prior to the coverage of WISE/NEOWISE, which explains their classification as {\it linear(-)} sources. The spectra of YSOs SPICY 31759  and 100587 show similarities to that of classical FUors and are therefore classified as this type of outbursts \citep[see ][]{2023Contreras_b,2024Guo_a}. The spectrum of SPICY 36590, taken close to the maximum point in the light curve, shows $^{12}$CO emission in spite of the high-amplitude and long-duration of the outburst. The magnetospheric accretion process might still be in control in this type of eruptive variable YSO \citep[][and Section \ref{ssec:peculiar}]{2021Guo}. 

\noindent \textit{\textbf{Linear($+$)}}: SPICY 11492 (L222\_1), 63130 (L222\_148), 15180 (Stim 1) and 29017 (VVVv631), are all part of previous publications from the VVV team \citep{2017Contreras,2021Guo,2024Guo_a,2024Lucas}. The four sources show long-term rises in their light curves, both at near- and mid-IR wavelengths. These long-term trends were noticed by \citet{2024Guo_a}, with SPICY 63130 also showing quasi-periodic ($P\sim378$~d) variability on top of the long-term rise. Only in SPICY 11492 the spectrum resembles that of classical FUors with strong $^{12}$CO absorption at near-IR wavelengths. SPICY 63130 and 15180 both show emission line spectra. Finally, SPICY 29017 shows $^{12}$CO emission and H$_{2}$O absorption. This combination, which has also been observed in the eruptive YSO V1647 Ori, could be explained by a hot inner disc, which generates the $^{12}$CO emission, and where water absorption is generated in a cooler location further out in the disk \citep{2020Guo}.

SPICY 99341 and 95397 are both classified as {\it linear(+)}, however, due to the gap between WISE and NEOWISE observations we cannot claim that these are examples of slow-rising outbursts. SPICY 99341 is source \#266 of the UKIDSS GPS high-amplitude variable star sample \citep{2017Lucas} as it shows $\Delta K\simeq4$~mag. The spectroscopic follow-up of the YSO yields an FUor classification \citep{2023Contreras_b}. SPICY 95397 has previously been classified as a mid-IR high-amplitude star \citep[source \#21,][]{2020Lucas}, but its spectrum shows strong $^{13}$CO absorption. The strength of these rovibrational bands depends on the abundance of $^{13}$C \citep{2023Poorta}. Therefore, these absorption bands are commonly found in the spectra of evolved stars, where chemical enrichment has led to $^{12}$C/$^{13}$C ratios of $<30$ \citep[see e.g.][]{2019Hinkle}. These ratios are lower than the one found in the interstellar medium (ISM), and also expected in YSOs, of $^{12}$C/$^{13}$C$\sim$89 \citep{2023Poorta}. Given the observed $^{13}$CO features, SPICY 95397 is classified as a contaminating evolved source by \citet{2023Contreras_b}.

\subsection{New Objects from IRTF/SpeX}

\begin{figure}
	\resizebox{\columnwidth}{!}{\includegraphics[angle=0]{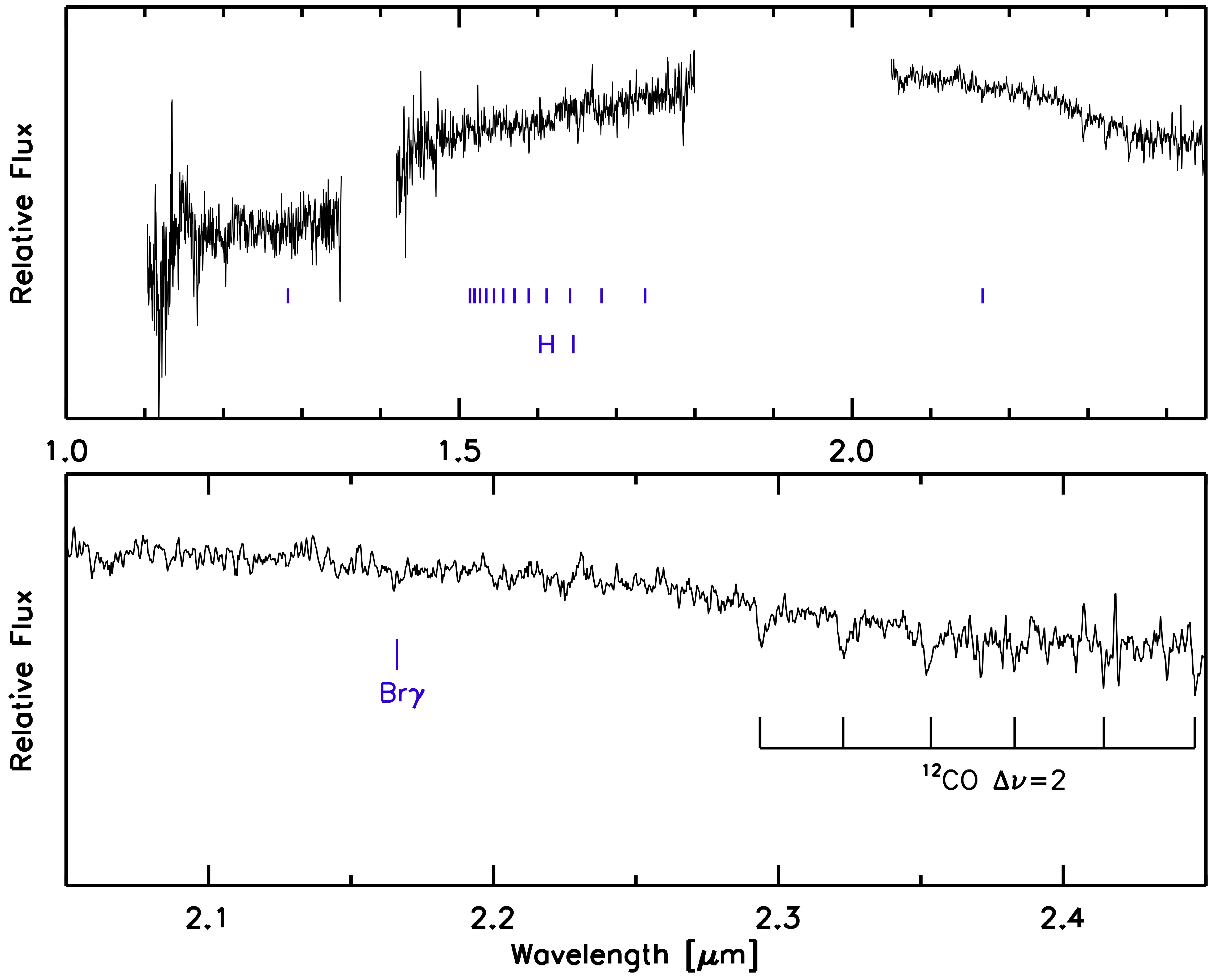}}\\
         \resizebox{\columnwidth}{!}{\includegraphics[angle=0]{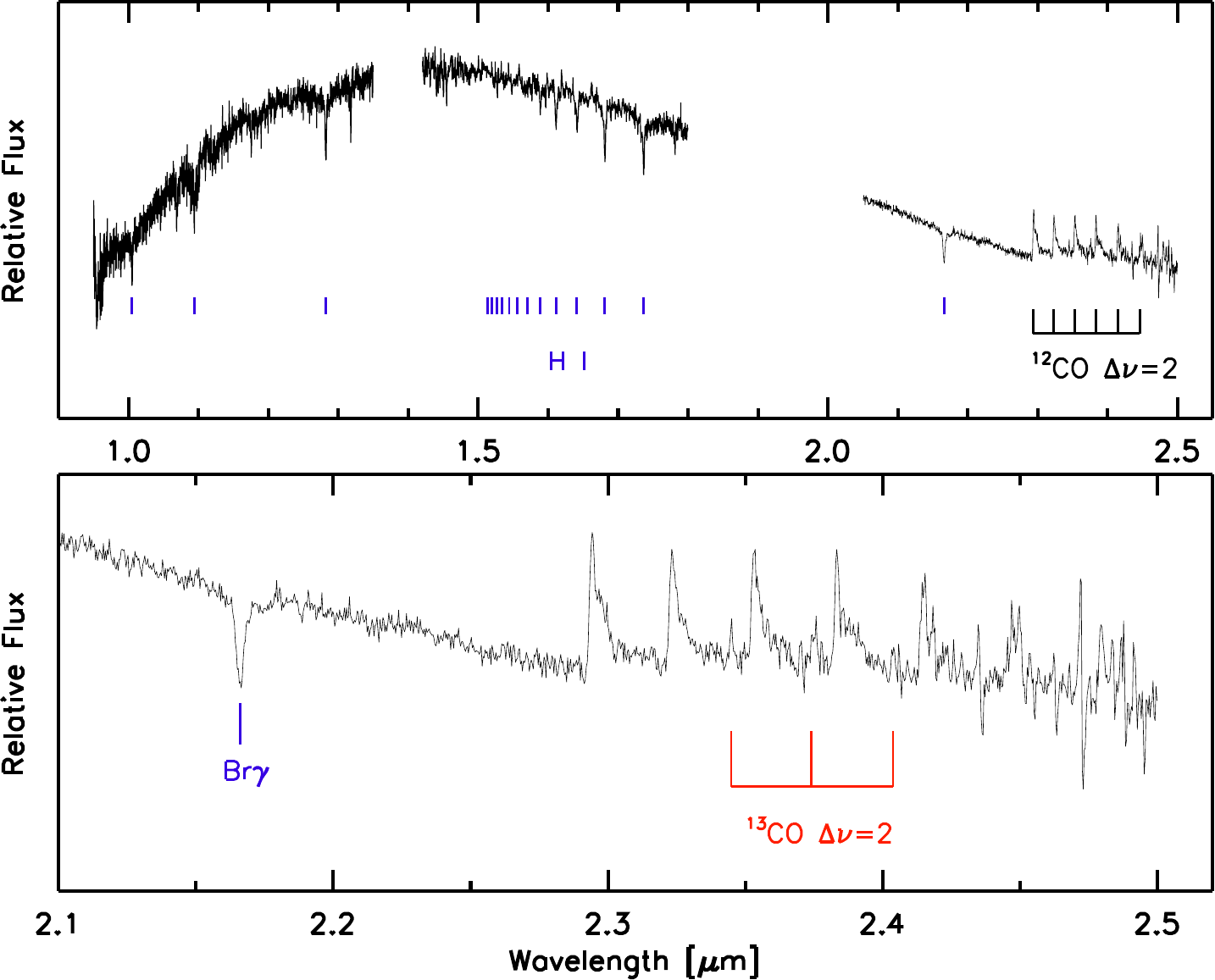}}
	 \caption{Near-IR spectra of SPICY 109102 (top two panels) and SPICY 111302 (bottom two panels). The location of H {\sc I} lines from the Paschen and Brackett series, as well as from $^{12}$CO $\nu=2-0$ rovibrational transitions are marked in both figures. For SPICY 111302 we also mark the location of  $^{13}$CO $\nu=2-0$ rovibrational transition lines.}
	 \label{fig:yso_irtf}
\end{figure}

SPICY 109102 and 111302 are both classified as {\it linear(-)} based on their WISE/NEOWISE light curves. The long-term behaviour is confirmed by {\it Spitzer} and 2MASS observations. The near-IR IRTF/SpeX spectra of both sources are shown in Fig. \ref{fig:yso_irtf}.

The near-IR spectrum of SPICY 111302 shows strong $^{12}$CO emission as well as absorption from several Hydrogen recombination lines (see Fig. \ref{fig:yso_irtf}). $^{12}$CO emission is generally observed in eruptive YSOs classified as EX Lupi-type \citep[e.g.][ and Section \ref{ssec:peculiar}]{2012Lorenzetti}. The presence of $^{13}$CO $\Delta\nu=2$ bandhead emission, however, is a strong indicator of chemical enrichment, and has been observed in the spectra of evolved stars \citep{2012Oksala,2020Kraus,2021Cochetti,2023Poorta}. The $^{13}$CO emission and strong \ion{H}{1} absorption are not a feature (to the best of our knowledge) of EX Lupi-type objects. In addition, the lack of \ion{Na}{1} emission is not common in the latter class of eruptive YSOs. The features are more likely to arise from a decretion disk, indicating that SPICY 111302 is an evolved star rather than a YSO.

The spectrum of SPICY 109102 is nearly featureless, except for the presence of strong $^{12}$CO absorption beyond 2.29 $\mu$m. We cannot detect any \ion{Ca}{1}, \ion{Na}{1} or water vapour absorption. The strength of $^{12}$CO compared with \ion{Na}{1}$+$\ion{Ca}{1} would be in line with a FUor classification \citep{2018Connelley}. However, the lack of an observed outburst would only allow for a FUor-like classification of the source.

\subsection{High resolution IGRINS spectra: FUor-like sources}\label{ssec:hr_abs}

\begin{figure*}
	\resizebox{\columnwidth}{!}{\includegraphics[angle=0]{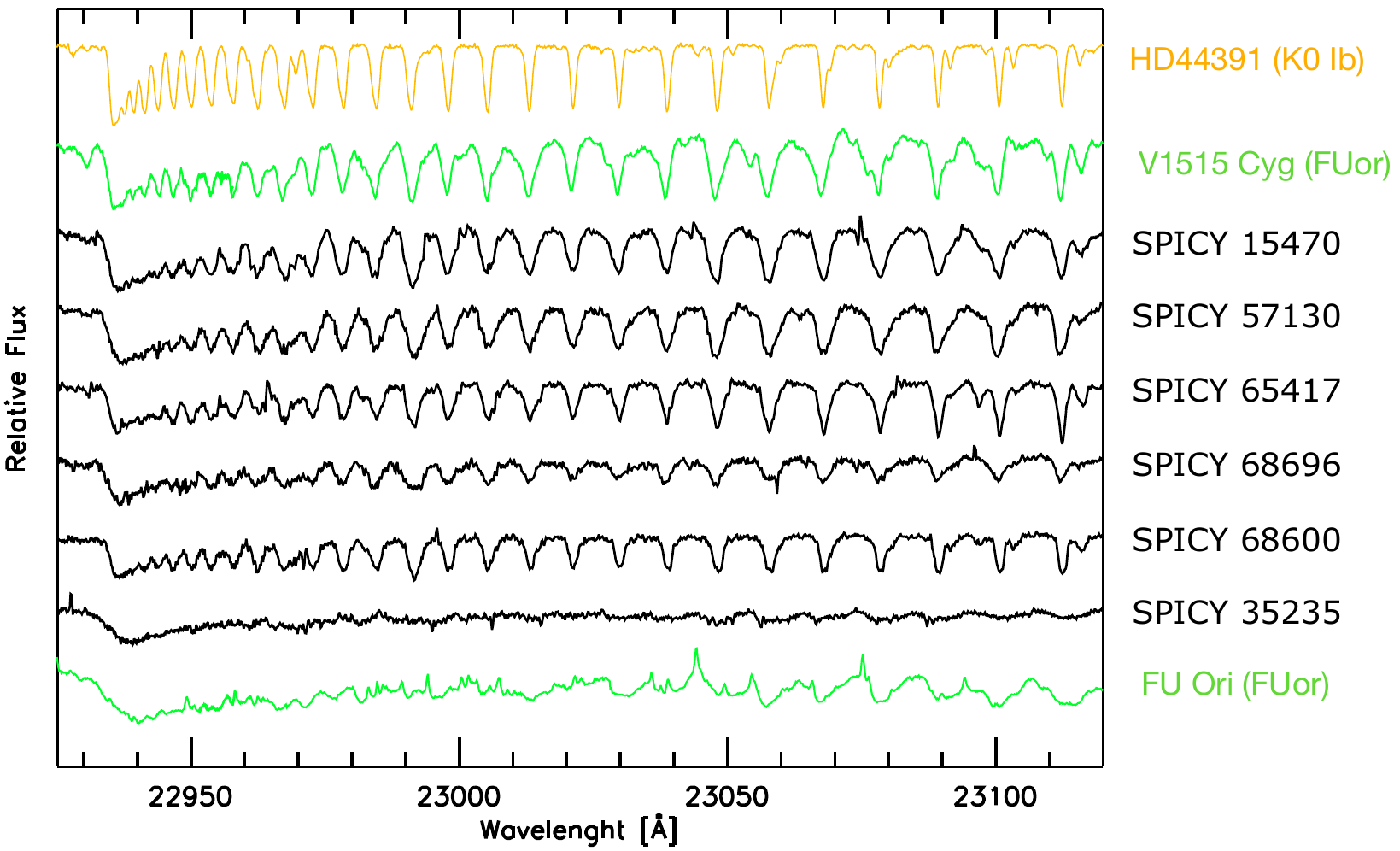}}
	\resizebox{\columnwidth}{!}{\includegraphics[angle=0]{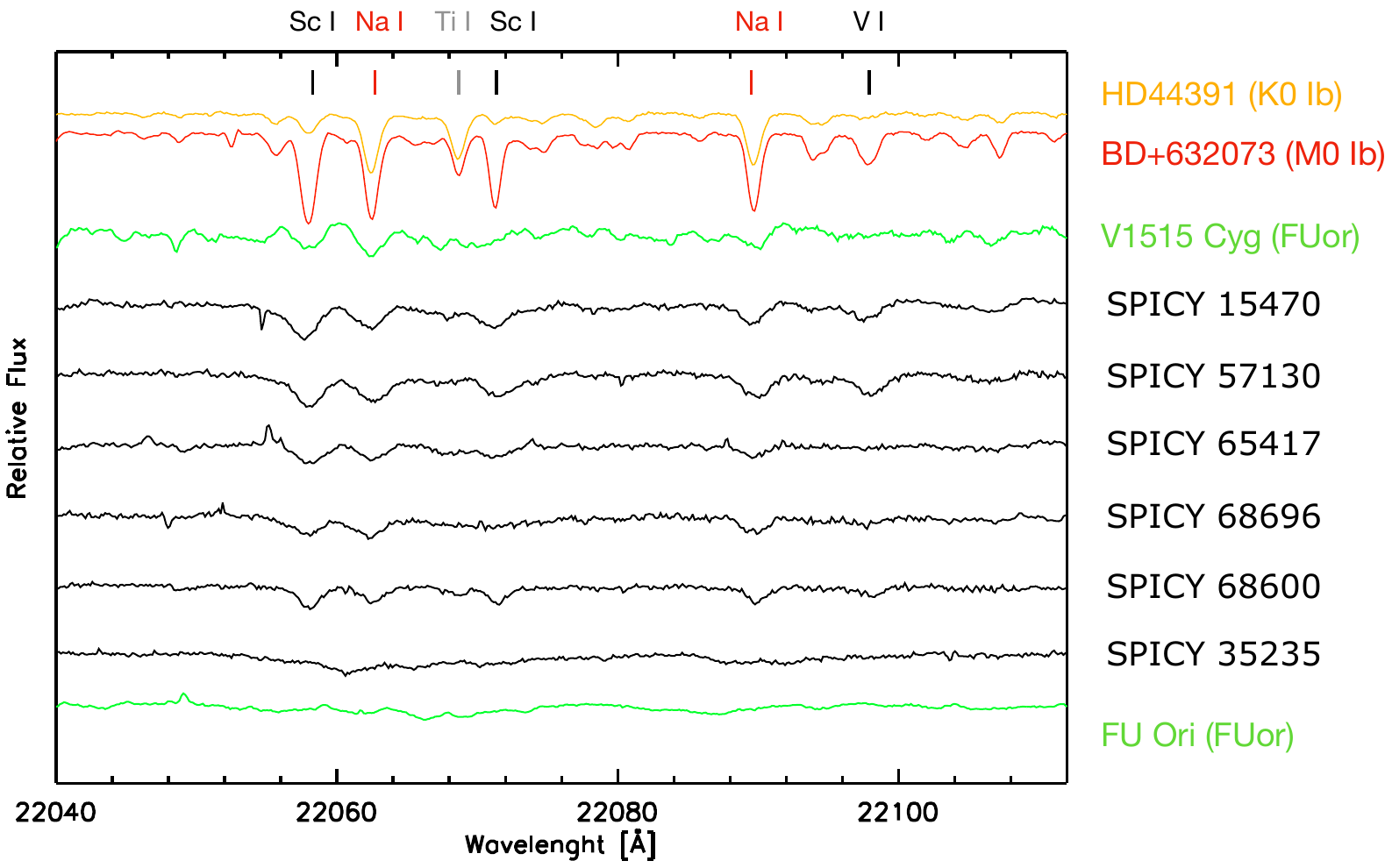}}
	 \caption{IGRINS spectra of six FUor candidates from our selection (solid black lines). The spectra is centered around the wavelength of the $^{12}$CO $\nu=2-0$ bandhead absorption.}
	 \label{fig:co1}
\end{figure*}

Figure \ref{fig:co1} shows the $^{12}$CO $\nu=2-0$ region of the spectra for six candidate FUors with IGRINS observations. Their spectra show strong and broad $^{12}$CO absorption that resembles that of the known FUors V1515 Cyg and FU Ori. In FUor outbursts, line broadening occurs due to disk rotation. V1515 Cyg is the classical FUor with the least broadened $^{12}$CO absorption profile, which is attributed to the lower inclination of the system, as compared with e.g. FU Ori \citep{2019Hillenbrand, 2022Szabo}. We note that the IGRINS spectra of SPICY YSOs 15470 and 57130 look very similar to that of V1515 Cyg, which could be evidence of low disk inclinations in these systems. 

The spectra around the \ion{Na}{1} absorption lines provide more evidence that the near-IR spectra of our sample arise from a disk. The six objects show absorption from \ion{Na}{1} (Fig.~\ref{fig:co1}). These features are not so evident in SPICY 35235, likely due to them being strongly broadened, similar to the spectrum of FU Ori itself. In five out of six objects (Fig.~\ref{fig:co1}) we observe absorption from \ion{Sc}{1} lines at 22057 and 22071 \AA, and for two objects (SPICY 15470 and SPICY 57130), we observe absorption from \ion{V}{1} at 22097 \AA. Comparison with V1515 Cyg shows that \ion{Sc}{1} absorption is also present in this classical FUor. The \ion{Sc}{1} and \ion{V}{1} lines are stronger in the spectra of M-type giant stars, as compared with K-type atmospheres. For example, \citet{2011Takagi}~shows that the ratio between the \ion{Sc}{1} and \ion{Na}{1} lines is sensitive to surface gravity, where objects with $\log~g<2$ show stronger absorption from \ion{Sc}{1}.

To provide further evidence of the disk nature of the absorption, the H- and K-band absorption of the six candidate FUors is fitted by convolving stellar templates from the IGRINS spectral library \citep[][]{2018Park} using a Keplerian disk profile to account for broadening \citep[see equation 1 in][]{2021Yoon}. The value of the maximum projected rotation velocity ranges between 7 and 110 km s$^{-1}$, and a veiling factor, $r$, is also introduced following \citet{2021Kidder} and \citet{2021Yoon}. We used $\chi^{2}$ minimisation to find the best fit. Following \citet{2015Lee} and \citet{2020Park}, 29 regions in the H and K-band are analyzed, which include absorption from, e.g., \ion{Fe}{1} ($15340$, $15670$, and $16490$~\AA), \ion{Al}{1} ($16755$~\AA) , \ion{Na}{1} ($22060,22090$~\AA), \ion{Ti}{1} ($21789$~\AA) and $^{12}$CO $\nu=2-0$ ($22935$~\AA).

\begin{figure*}
\centering
	\resizebox{1\columnwidth}{!}{\includegraphics[angle=0]{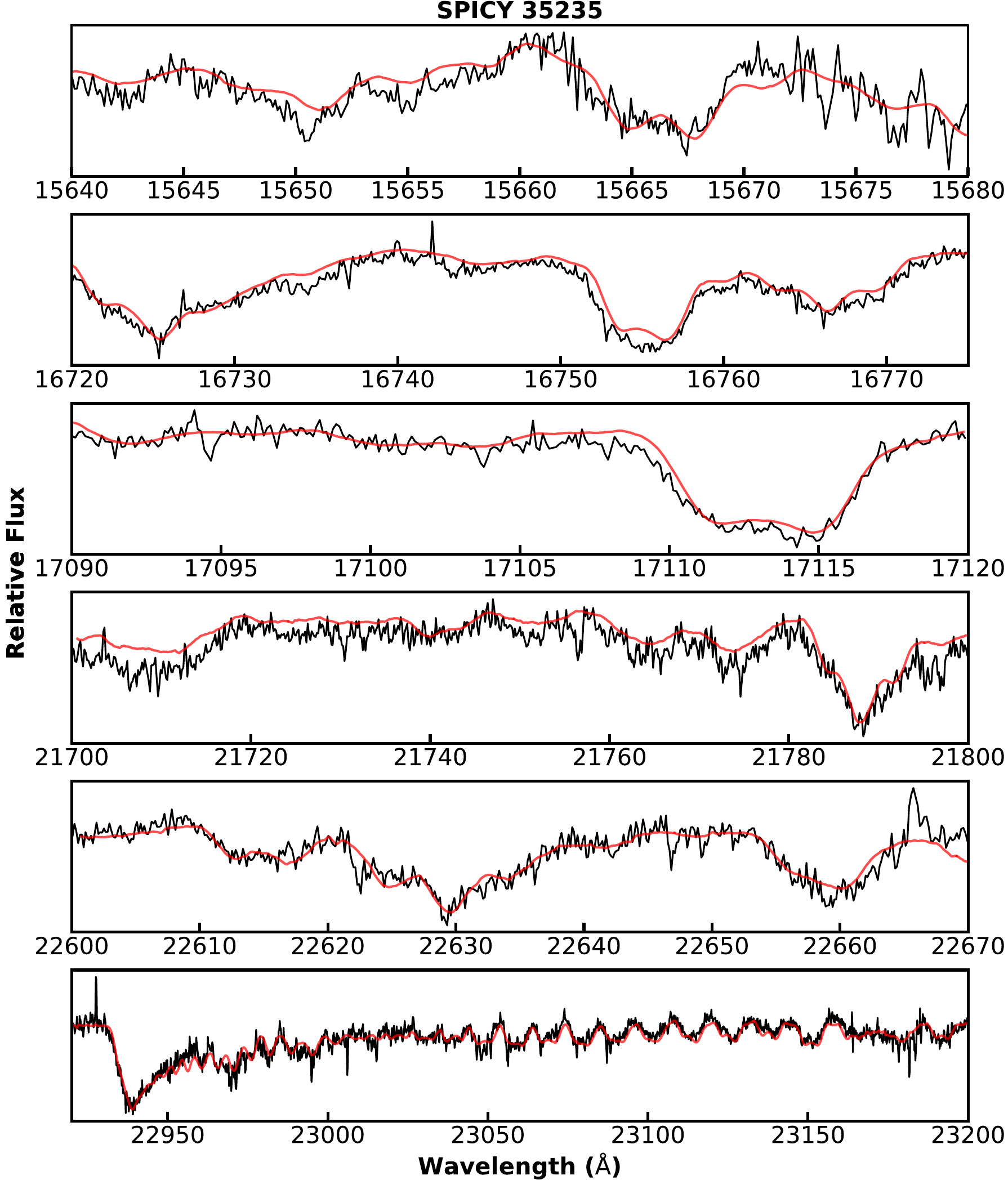}}
	\resizebox{1\columnwidth}{!}{\includegraphics[angle=0]{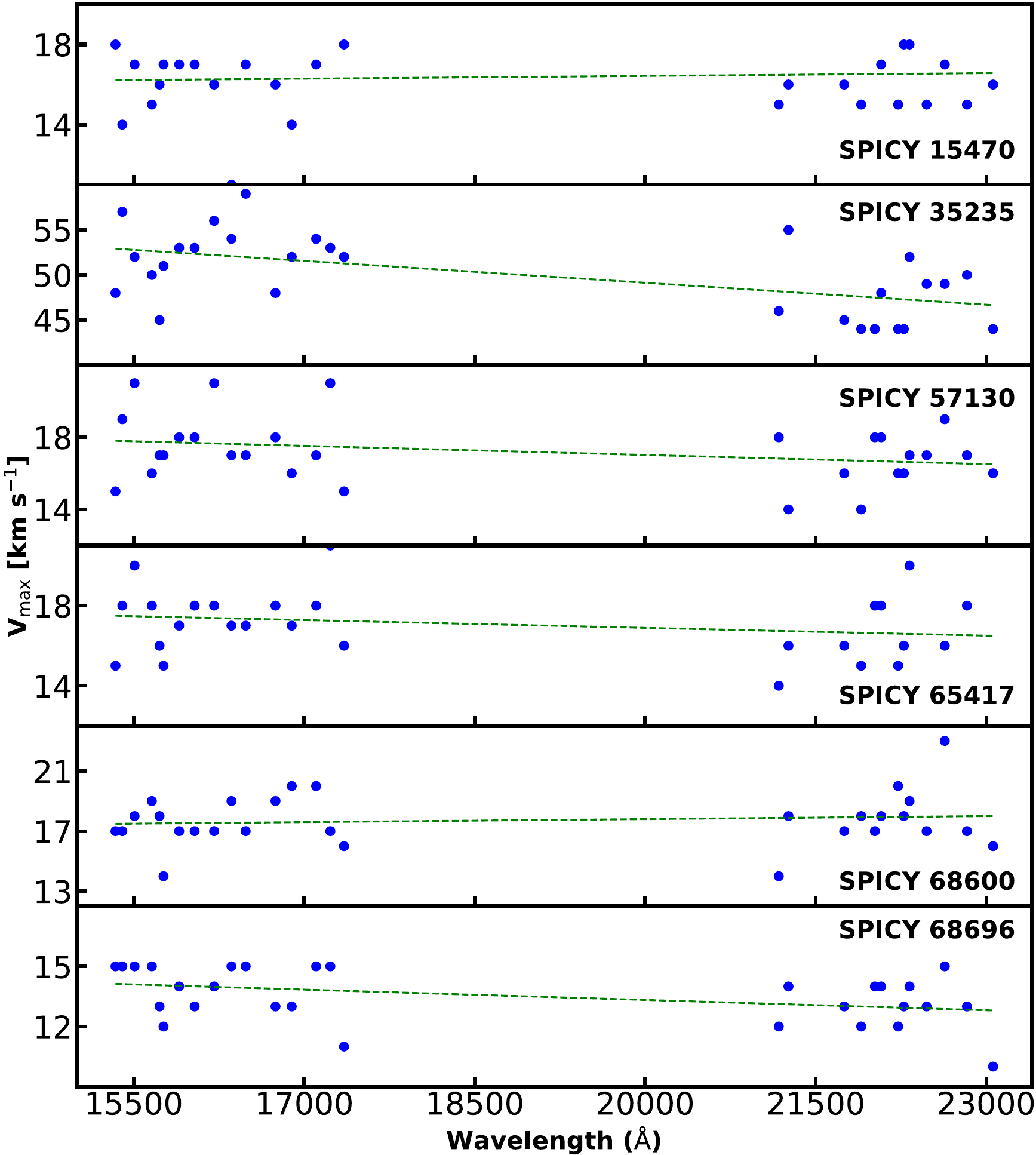}}
	 \caption{(left) IGRINS spectra of SPICY 35235 (solid black lines) and the fits using templates from IGRINS spectral library (red), as described in the main text. (right) $v \sin i$ vs wavelength for the 29 regions used to fit the data. The green dashed lines mark linear fits to these data.}
	 \label{fig:co2}
\end{figure*}

The analysis of these 29 spectral regions in the H- and K-band data for the six YSOs, shows that G8-M3 templates with $v \sin i$ between 12 and 55~km s$^{-1}$ provide good fits to the spectra (see Fig.~\ref{fig:co2} for example regions in the spectrum of SPICY 35235). The expectation in a Keplerian disk model is that the rotational velocity of the disk decreases with wavelength. This is expected because warmer regions near the protostar rotate faster than cooler outer regions \citep{2021Yoon}. The increase in $v \sin i$ has been observed in FUor disks, being more noticeable when comparing optical to infrared wavelengths \citep[e.g.][]{2007Zhu,2015Lee}. In four sources we observe a decrease in the maximum projected rotational velocity as we move towards longer wavelengths, with SPICY 35235 showing the most dramatic change. In two sources, SPICY 15470 and SPICY 68696, the velocities appear to remain constant, and given the lack of optical data, we cannot confirm any decrease towards longer wavelengths. A low value of the inclination in these systems could explain the relatively constant velocities in the 1.5-2.4 $\mu$m range.

The lack of an outburst observation in all of these sources still raises some concern regarding possible contamination. In Section \ref{ssec:contamination}, we have stated that giant stars in D-type symbiotic binary systems show large stellar rotation velocities that could lead to line broadening. These velocities are similar to the ones found for the six sources with IGRINS data. However, \citet{2008Corradi} shows that D-type symbiotic stars also have large r-H$\alpha$ excesses (see their figure 1). Three of our sources, are found in the VPHAS catalogue \citep{c2016Drew}. The colours of SPICY 57130 ($r-i=1.3$, $r-H\alpha=0.7$), SPICY 68600 ($r-i=2.1$, $r-H\alpha=0.6$) and SPICY 68696 ($r-i=2.1$, $r-H\alpha=0.6$) are inconsistent with a r-H$\alpha$ excess and therefore disagree with a D-type symbiotic classification. In addition our spectra lack the strong HI and HeI emission that is commonly observed in symbiotic stars \citep{2000Belczynski}.

Therefore, the evidence presented above for the spectral regions in the H- and K-bands, i.e. broadening consistent with a Keplerian rotation model, favours the accretion disk scenario in our objects.

\subsection{High resolution IGRINS spectra: Peculiar outbursts}\label{ssec:peculiar}

\begin{figure*}
	\resizebox{2\columnwidth}{!}{\includegraphics[angle=0]{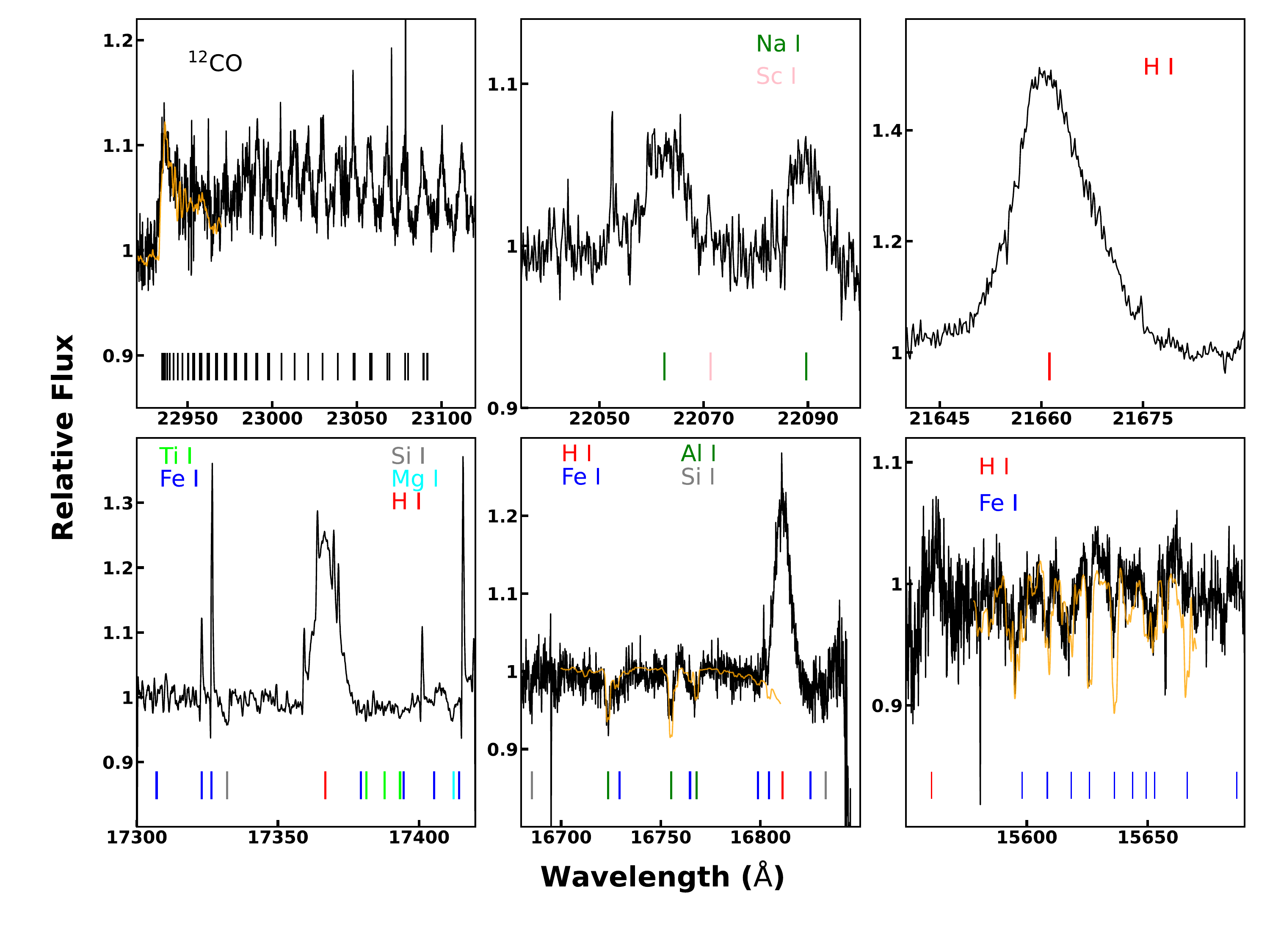}}
	 \caption{IGRINS spectra covering regions in the H and K bands for SPICY 21349. In the plot, we mark the wavelengths of transition lines from several elements. The orange solid line shows the result of convolving stellar templates with a Keplerian rotation model.}
	 \label{fig:emission}
\end{figure*}

Two sources, SPICY 21349 and SPICY 42901, show spectra featuring emission in Hydrogen Br$\gamma$ (2.16 $\mu$m), \ion{Na}{1} (2.26 $\mu$m), and $^{12}$CO $\Delta \nu=2-0$ (2.2935 $\mu$m) (see Fig. \ref{fig:emission}). These characteristics were usually associated with EX Lupi-type or V1647-Ori like outbursts, which display outburst durations that are shorter than the typical timescales of FUor outbursts. However, there is growing evidence that long-term outbursts can also display emission line spectra.

The known eruptive YSO SVS13 went into outburst between 1988 and 1990 and has remained bright ever since, with the spectrum of the object being dominated by emission lines during the outburst \citep{1991Eisloeffel}. YSOs VVVv270 and VVVv631 \citep{2017Contreras} also show long-duration outbursts while displaying emission line spectra during various epochs collected at bright state \citep{2020Guo}. In fact the majority of eruptive YSOs that show long-duration outbursts in the sample of \citet{2021Guo} may be controlled by the magnetospheric accretion process, as indicated by the Br$\gamma$ emission. Therefore the characteristics of the spectrum of SPICY YSOs 21349 and 42091 do not contradict the expectations that these objects are going through long-term outbursts.

\begin{figure*}
 \resizebox{2\columnwidth}{!}{\includegraphics[angle=0]{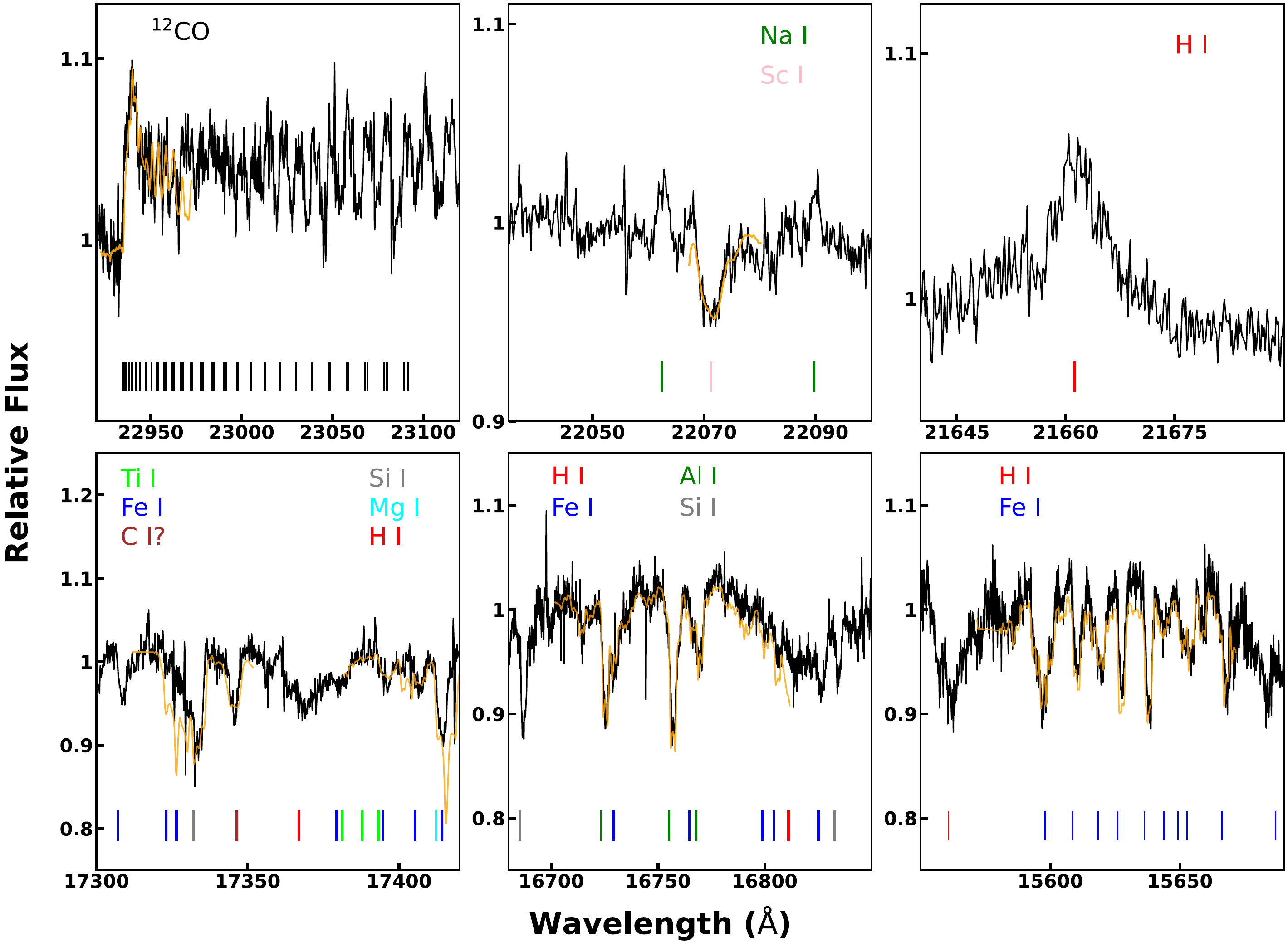}}
	 \caption{IGRINS spectra covering regions in the H and K bands for SPICY 42901. In the plot, we mark the wavelengths of transition lines from several elements. The orange solid line shows the result of convolving stellar templates with a Keplerian rotation model.}
	 \label{fig:emission2}
\end{figure*}

To study the $^{12}$CO $\nu=2-0$ emission in both YSOs, we used a similar method as in Section \ref{ssec:hr_abs}, but multiplying the stellar templates by $-1$ to turn the CO features into emission. We find that these features are broadened with $v\sin i=27$ km s$^{-1}$ and $v\sin i=39$ km s$^{-1}$ for SPICY 21349 and 42901, respectively, consistent with broadening expected from Keplerian rotation \citep{1996Najita}. The strength of the \ion{Na}{1} and Br$\gamma$ emission varies depending on the YSO, with SPICY 21349 showing stronger emission of these spectral features. Several emission lines from the Hydrogen Brackett series are also present in the H-band spectra of the source (see Fig.~\ref{fig:emission}). Interestingly, the H-band spectrum of SPICY 42901 lacks any further emission from Hydrogen lines and is instead dominated by absorption from e.g. \ion{Fe}{1} and \ion{Al}{1} (Fig.~\ref{fig:emission2}). In addition, \ion{Sc}{1} is seen as an absorption line at 2.207 $\mu$m. These absorption lines appear broader than expected from a stellar photosphere. Some absorption lines also appear broadened in the spectrum of SPICY 21349, but it is less clear in this source. This combination of broad absorption lines and emission lines, strengthens the scenario where the flux arises from multiple locations in an accretion disk.

\section{On the timescales and frequencies of FUor outbursts}\label{sec:timescales}

Based on the results from YSOs with spectroscopic data, we find 18 objects that are classified as eruptive variable YSOs, with 9 of these corresponding to new additions to the variable class (7 FUor-like, 2 Peculiar). Only two sources are found to be evolved stars (one AGB and one Be star).

Given these results, the method outlined in this work is found to be highly successful in finding eruptive YSOs in the long-term VVV and WISE/NEOWISE data. By assuming that the majority of objects in our sample are true eruptive YSOs, we can provide some insights into the rising and decay timescales of FUor outbursts, as well as reconciling the number of objects in our sample with the estimated frequencies of FUor outbursts.

\subsection{Outburst timescales}\label{ssec:otime}

Under the assumption that most of the 717 {\it linear} YSOs are true eruptive YSOs, then the number of sources classified into the {\it linear(-)} and {\it linear(+)} categories can, tentatively, provide some insight into the rising timescales of FUor outbursts. From the sample of 717 YSOs classified as {\it linear} we find a ratio of rising ({\it linear(+)}) to fading ({\it linear(-)}) objects of 171/526, or $\sim$0.32. 

\begin{figure}
	\resizebox{\columnwidth}{!}{\includegraphics[angle=0]{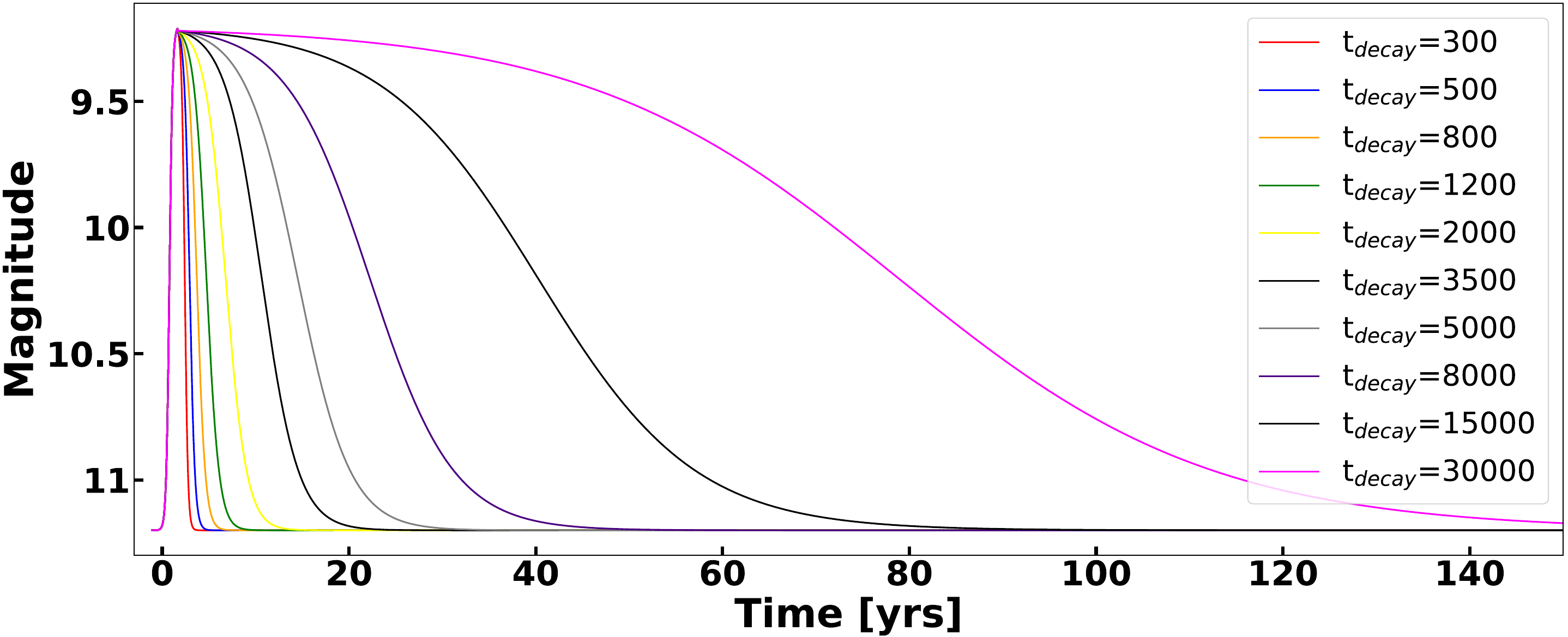}}\\
	\resizebox{\columnwidth}{!}{\includegraphics[angle=0]{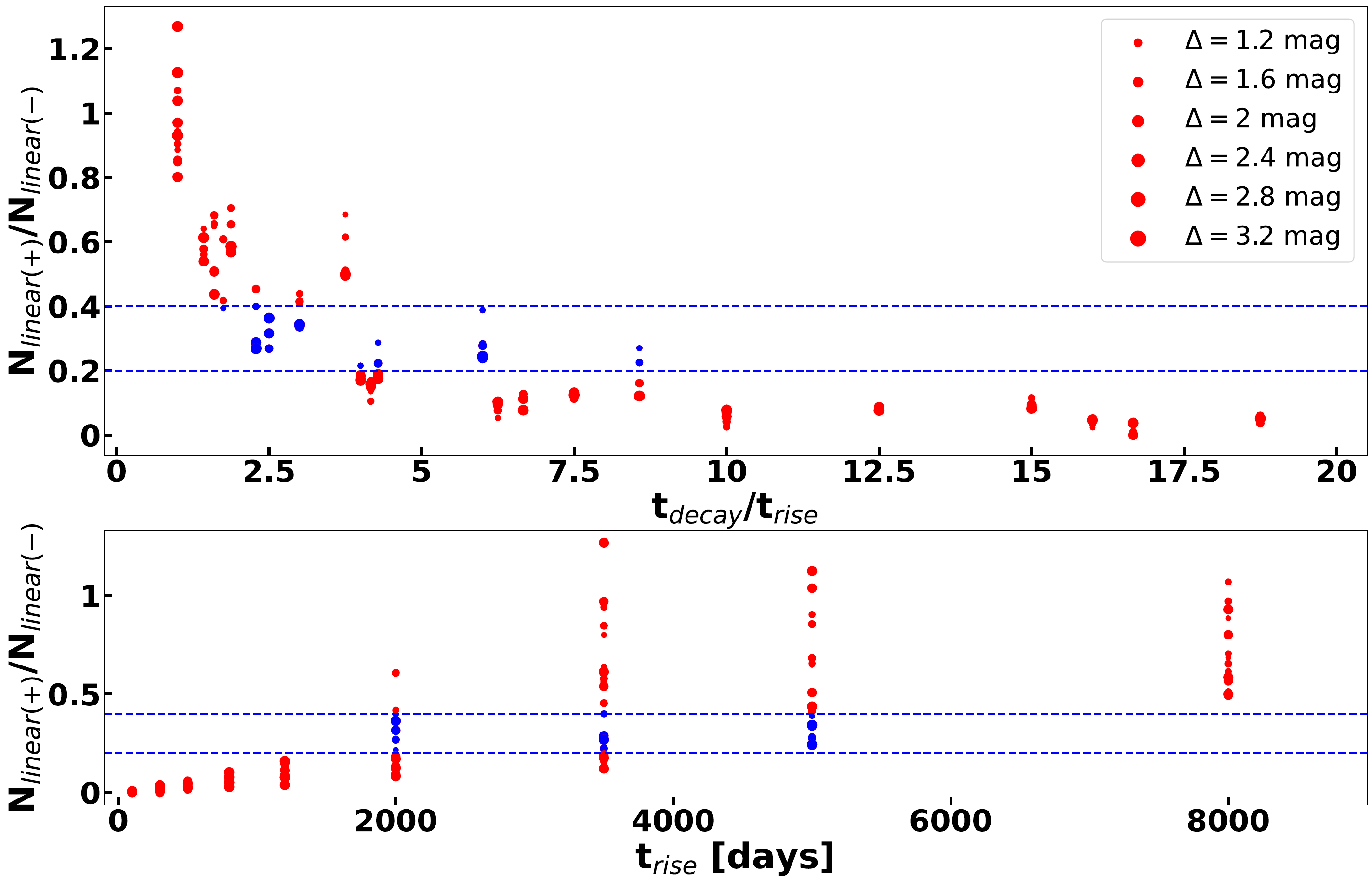}}

	 \caption{(Top) Synthetic FUor light curves for $t_{\rm rise}=300$ days, amplitude of 2 mag, and for different decay timescales of the outburst. (Middle, Bottom) Ratio of rising ({\it linear(+)}) to fading ({\it linear(-)}) objects as a function of $t_{\rm decay}/t_{\rm rise}$ (middle plot) and $t_{\rm rise}$ (bottom plot). In both figures, the size of circles increases with the amplitude of the outburst. Blue circles mark the results that yield {\it linear(+)} to {\it linear(-)} between 0.2 and 0.4. These regions are also marked by dashed blue lines.}
	 \label{fig:times}
\end{figure}

To quantify the significance of the ratio in terms of rising and decay timescales of outbursts, we constructed a set of synthetic light curves that simulate FUor outbursts. These were built by combining two sigmoid functions, one that simulates the rise and another to simulate the decline of the outburst. The starting time of the latter function is set as the final time of the rise function.  The rise timescales, $t_{\rm rise}$ (or the time that it takes to reach 90\% of the peak brightness), are set at 100, 300, 500, 800, 1200, 2000, 3500, 5000 and 8000 days. The decay timescales, $t_{\rm decay}$, are set at the same values, but with the condition that  $t_{\rm rise}\leq t_{\rm decay}$. For all values of $t_{\rm rise}$ we also simulate light curves with $t_{\rm decay}=15000$~and 30000 days. The amplitude of the bursts are also varied and are set as $\Delta=1.2$, 1.6, 2, 2.4, 2.8 and 3.2 mag. The choice of parameters is based on previous observations of FUors. Amplitudes larger than 1 mag are expected in the mid-IR for accretion-driven variability \citep{2013Scholz,2022Rodriguez}. The rise timescales in the FUors discovered in the VVV survey range between 100 and 2500 days \citep{2024Guo_a}.We also allow for longer values of $t_{\rm rise}$ as some of the theoretical mid-IR outbursts by \citet{2023Cleaver} can show rises lasting longer than 20 years, and the mid-IR rising timescales are usually longer than the near-IR timescales, such as in Gaia17bpi \citep{2018Hillenbrand} and L222\_78 \citep{2024Guo_b}. Finally, known FUors show shorter values of $t_{\rm rise}$ compared with the time they spend at bright state \citep[see e.g.][]{2014Audard}. Examples of simulated light curves are shown in Figure \ref{fig:times}. 

For given values of $t_{\rm rise}$, $t_{\rm decay}$ and amplitude, we create a synthetic light curve. Then we follow the steps below:

\begin{itemize}
    \item[(a)] Select 20 points that follow the observation cadence of WISE/NEOWISE data. The starting point of the observations is selected randomly in the synthetic light curve;
    \item[(b)] Apply a random measurement error with a maximum value of 0.08 mag;
    \item[(c)] Classify the new WISE/NEOWISE light curve using the method of Section \ref{sec:lc_class};
    \item[(d)] Steps (a) to (c) are repeated 2000 times.
\end{itemize}

We note that $\sim$20\% of the SPICY YSOs are found at either the bright or faint end of the mid-IR magnitude distribution, and therefore show larger errors ($>0.1$~mag). To account for the effect of larger errors in the classification process, in $\sim$20\% of the repetitions (selected randomly) we assign random errors with values between 0.1 and 0.2 mag in step (b).

Figure \ref{fig:times} shows the ratio of light curves classified as {\it linear(+)} to those classified as {\it linear(-)} as a function of $t_{\rm decay}/t_{\rm rise}$ (middle plot) and $t_{\rm rise}$ (bottom plot). The figure shows that to get {\it linear (+)} to {\it linear (-)} ratios similar to our results, we require the existence of outbursts with $2000<t_{\rm rise}<5000$ days and for $t_{\rm decay}$ to lie between about twice and five times the value of $t_{\rm rise}$. Light curves with equal values for the rise and decay of the outburst tend to produce an equal number of {\it linear (+)} and {\it linear (-)} objects, while larger values of $t_{\rm decay}$ compared with $t_{\rm rise}$ tend to overproduce  {\it linear(-)} objects.

\begin{figure}
 \resizebox{\columnwidth}{!}{\includegraphics[angle=0]{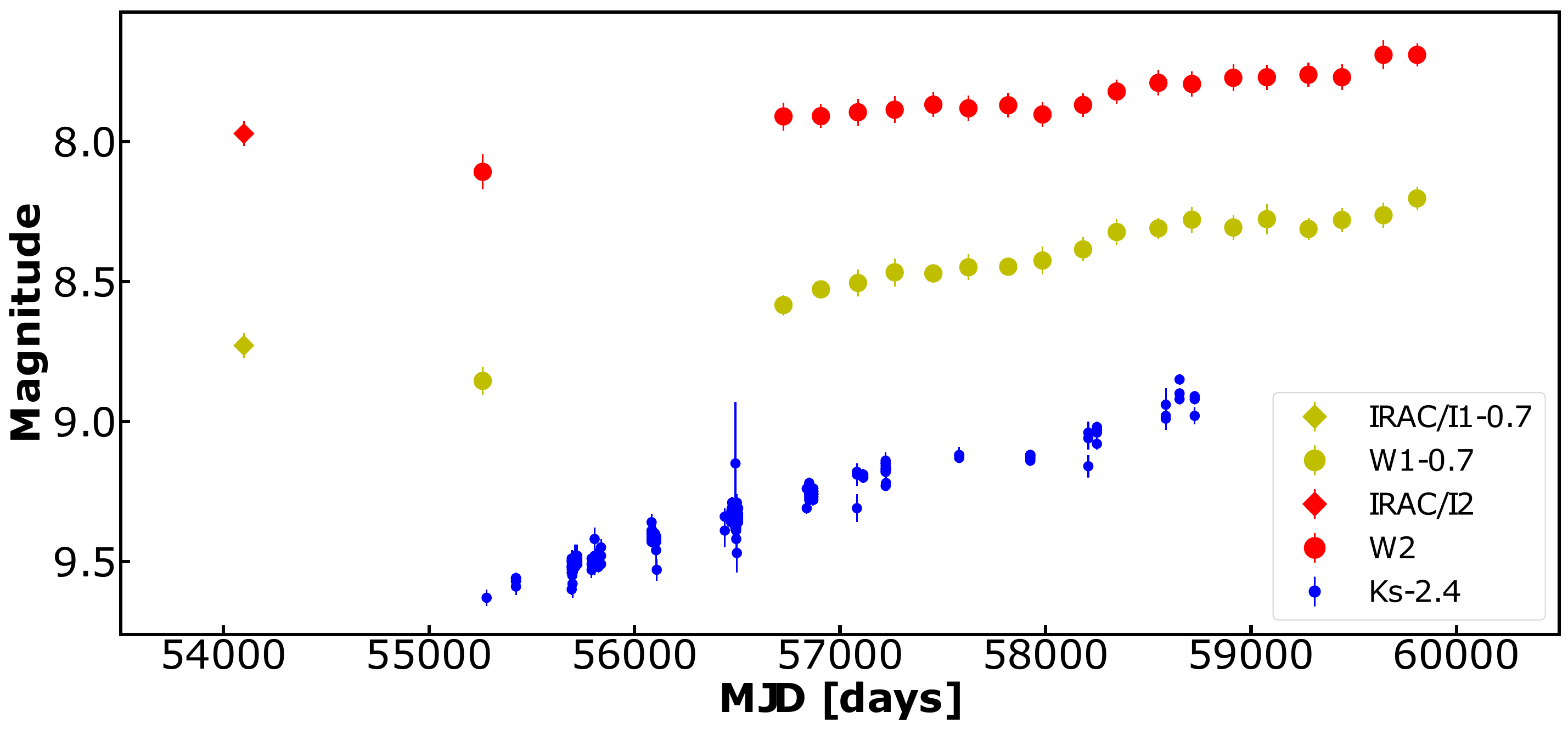}}
	 \caption{Near- to mid-IR light curve of SPICY 41259.}
	 \label{fig:41259}
\end{figure}

Examples of known FUors with $t_{\rm rise}>2000$~d are rare. Recently, \citet{2024Ashraf} identified the YSO SSTgbs J21470601$+$4739394 as an FUor outburst of a $\simeq$0.2M$_{\odot}$ star. The mid-IR light curve of the source shows a slow rise over the 10-year monitoring of WISE/NEOWISE observations. The classical FUor V1515 Cyg took 30 years to reach its peak brightness \citep{1977Herbig,2022Szabo}. The existence of decades-long rises is interesting, as according to the models by \citet{2023Cleaver}, such long mid-IR rises are expected in the outbursts reaching the largest accretion rates. In this sense, it is important to continue monitoring these sources (such as SPICY 41259, see Fig.~\ref{fig:41259}).

The values of $t_{\rm rise}$ and $t_{\rm decay}/t_{\rm rise}$ found from our sample can also provide a ``typical'' timescale of an FUor outburst. If we assume that $5000>t_{\rm rise}>2000$~d and $6>t_{\rm decay}/t_{\rm rise}>2.5$ then we find outburst durations that vary from 19 to 95 years, with a mean of 48 years.

\subsection{Comparison with the frequency of FUor outbursts}

The frequency of outbursts at different stages of YSO evolution has been measured through the detection of FUors in optical, near- and mid-IR surveys \citep{2013Scholz,2019Contreras,2021Park, 2022Zakri} as well as through analysis of the chemical imprint left by outbursts \citep{2015Jorgensen,2019Hsieh}. These methods all roughly agree on the increase in the recurrence timescales, $\tau$, as the systems evolve. The frequency of bursts is estimated as once every $\tau=112000$ yr at the Class II stage \citep{2019Contreras}, $\tau=1750-10000$ yr at the Class I stage \citep{2013Scholz,2019Hsieh,2024Contreras}  and $\tau=490-2000$ yr at the Class 0 stage \citep{2019Fischer,2021Park,2022Zakri}.

We can use the estimates of $\tau$ presented above to give a rough estimate of the mean lifetime of FUor outbursts. Setting $\tau=2000, 10000$ and $112000$~yrs for the Class I, Flat-Spectrum and Class II stages, respectively, we expect to observe $5\times10^{-4}$, $10^{-4}$ and 8.9$\times10^{-6}$ outbursts per star per year at the different stages. Given the sample sizes at each evolutionary stage (see Section \ref{ssec:sample}), we then determine that we should observe roughly 7.9, 2.4 and 0.5 new outbursts per year in Class I, Flat-Spectrum and Class II stages, respectively, or $\sim11$ outbursts per year in the overall SPICY sample. 

Given the above yearly rate of new burst detection for the SPICY sample estimated using literature-based recurrence timescales, along with the 717 monitored on-going FUors (assuming that the 717 {\it linear} sources are indeed FUor outbursts), we estimate that the typical time that each FUor source must spend in an observable burst mode is $717/11\simeq65$ years. 

This number is affected upward by contamination and downward by the fact that we may have missed some outbursts. The spectroscopic sample shows a 10\% contamination from evolved sources. This sample is biased towards the brightest end of the distribution of candidate FUors, and therefore the 10\% may be a lower limit of the contamination level. if we assume that contamination can reach as high as 3 times this level, then the number of candidates reduces to $\sim$500 sources. A rough estimate of the number of outbursts we are missing can be derived from the number of times a ``typical'' synthetic light curve of an outburst would not be classified as a linear source. Based on the results from the previous section, we create a synthetic light curve with an amplitude of 2.4 mag, $t_{\rm rise}=2000$~d and $t_{\rm decay}/t_{\rm rise}=6.5$. We then followed steps (a) to (d) from the simulations of the previous section. We find that this outburst is classified as linear in 59\% of the 2000 repetitions. This would imply that the number of outbursts in our sample (after considering the 10\% lower limit of the contamination) could be as high as $\sim$1100 sources. 

The outbursting FUor range of $500<N_{\rm events}<1100$, implies a mean burst lifetime between 45 and 100 years. Although there are many caveats that affect how we realize this number, the range of values for the mean burst lifetime are fairly consistent with the ``typical'' duration of outbursts found above in Section \ref{ssec:otime}, as well as with the longest baselines we are able to retrieve from the SuperCOSMOS survey in Section \ref{sssec:light_curves}.

Theoretical models that study the processes that lead to an outburst predict a wide range of durations \citep[see figure 7 in ][]{2014Audard}. Stellar flybys can lead to outbursts lasting decades to centuries \citep{2022Borchert}. Planet-induced accretion events, and sudden triggering of magneto-rotational instabilities (MRI) in the inner disk can lead to durations typically lower than 60 years \citep{2004Lodato,2023Cleaver}. In general, gravitational instabilities (GI) that lead to disk fragmentation or that trigger MRI, predict the largest durations \citep[up to 500 years, ][]{2009Vorobyov,2020Kadam,2023Cleaver}. Some outbursts that occur due to disk fragmentation at the end of the embedded phase/early Class II stage, also show durations of a few decades \citep{2009Vorobyov}.

The mean-burst values derived from our analysis are inconsistent with the centuries-long timescales predicted by some of the models. However, the instabilities that lead to such long-timescales, are triggered during stages of large mass infall rates from the envelope. These large rates are associated with the deeply embedded Class 0 stage of YSO evolution. The SPICY catalogue is dominated by Class II and flat-spectrum YSOs, therefore the mean-burst values derived above may indicate that GI is not the dominant mechanism leading to outbursts in this sample.

\section{Summary}\label{sec:summary}

In this work we have attempted to discover YSOs that, 1) have gone into outburst prior to the start of modern monitoring or 2) show long-term slow-rising outbursts. For this we have used several properties of accretion-related outbursts. Both type of outbursts should be well fitted by linear models as they are either slowly declining after outburst \citep{2000Kenyon} or slowly rising towards peak brightness \citep{2024Ashraf}. Further, the YSOs that go into outburst should reach a similar location in mid-IR CMDs, independent of the mass of the central star \citep{2022Liu}. Finally, the near-IR spectra of these sources should show the characteristics of FUors, e.g. $^{12}$CO absorption broadened by disk rotation \citep{2018Connelley}.

We analysed the WISE/NEOWISE mid-IR light curves of 99702 YSOs arising from the SPICY catalog \citep{2020Kuhn}. Classification of their light curves using the methods from \citet{2021Park} allowed us to find 717 YSOs with light curves that are well fitted by a linear model, 526 with a negative slope and 191 with a positive slope. Using additional data from near-IR (VVV) data, as well as the location of these objects in W1 versus W1$-$W2 CMDs, we are able to determine candidates with the higher probability of being FUors.

Based on the results from YSOs with spectroscopic data (data obtained by our group, as well as data from the literature), the method outlined in this work is found to be highly successful in finding eruptive YSOs in the long-term WISE/NEOWISE data. We confirm that 18 out of 20 objects are eruptive variable YSOs. In only two objects we find spectroscopic evidence that points to an evolved nature of these sources. The eruptive variable YSOs can be divided into four bonafide FUors (as a high-amplitude outburst has been observed in the past), eight FUor-like sources (no outburst has been recorded) and 6 Peculiar sources (show a mix of characteristics from the known subclasses of eruptive YSOs). Nine of the YSOs analyzed in this work correspond to new additions to the variable class (7 FUor-like, 2 Peculiar). For a more detailed discussion of the classes of eruptive variable YSOs see \citet{2014Audard,2018Connelley,2023Contreras_a}.

 The spectra of 12 objects show a lack of emission lines and have strong absorption from $^{12}$CO. In six cases the FUOr classification arises from mid-resolution spectra \citep{2010Connelley,2023Contreras_b,2024Guo_a}  and our mid-IR light curves support this classification. In six sources we have high-resolution IGRINS spectra ($R\sim45000$) that show broadened absorption that can be fitted by convolving stellar templates with a Keplerian disk rotation profile. We also observe an increase in velocity with wavelength that is expected in the accretion disk of FUors. Contamination from D-type symbiotic sources is discarded as our sample does not show the H$\alpha$ excess (estimated from broadband photometry) that is expected in these evolved systems.

In six sources we find $^{12}$CO, Br$\gamma$ and \ion{Na}{1} in emission. These features are generally not attributed to the more extreme and longer duration outbursts (a.k.a FUors), which could argue against these sources being in a high-accretion state from a centuries-long outburst. However, there are an increasing number of eruptive YSOs that show long duration outbursts and an emission line spectrum during high state \citep{1991Eisloeffel,2017Contreras,2021Guo}. In these cases magnethospheric accretion would still be in control even at large accretion rates \citep{2021Guo}. The latter could be due to these YSOs having larger masses of the central star \citep{1991Calvet,2022Liu}. The mechanism driving the variability is likely still a large disk instability.

In the particular case of  SPICY 42901 we also observe that the H-band spectrum is dominated by broad absorption of \ion{Fe}{1} and \ion{Al}{1} and lack of Hydrogen emission. It is possible that the multiple components arise from different regions of the disk, with the hot midplane dominating the emission in the inner regions (leading to absorption in the H band), while CO emission arises from regions where the atmosphere of the disk has a larger temperature than the midplane. However, we could also be observing the combined flux in a binary system. For example, Z CMa is a binary system comprised of a Herbig Ae/Be star and an FUor, with the $^{12}$CO emission arising from the higher-mass object and the FUor dominating the flux at shorter wavelengths \citep[see e.g.][]{1993Whitney,2013Hinkley,2017Bonnefoy}.

The overall number of {\it linear(+)}  and {\it linear(-)} objects provide some insights into the timescales involved in accretion-driven outbursts. Firstly, we performed a Monte Carlo simulation of synthetic light curves of various amplitudes, rising and decay timescales to estimate the parameters that would lead to the observed ratio of {\it linear(+)} to {\it linear(-)} objects in our sample. The simulation yields that outburst decay times must be at least 2.5 times longer than the rise times. In addition, a population of outbursts with rise timescales between 2000 and 5000 days must exist to get our observed number of linear(+) YSOs. Using these results we also estimate that a ``typical'' outburst can last between 19 and 95 years, with a mean duration of 48 years. Secondly, we used previously measured values of the frequency of outbursts and determine that to detect between 500 and 1100 outbursts in the SPICY sample, FUor outbursts must have a mean burst lifetime of between 45 and 100 years. Both analyses yield consistent results in terms of outburst duration, and both agree with the longest baselines we are able to retrieve for our spectroscopic sample.

Future observations of the sample of 717 YSOs presented in this paper has the potential to uncover a large number of YSO outbursts. 

\vspace{2mm}

CCP was supported by the National Research Foundation of Korea (NRF) grant funded by the Korean government (MEST) (No. 2019R1A6A1A10073437). DJ is supported by NRC Canada and by an NSERC Discovery Grant. This work was supported by the New Faculty Startup Fund from Seoul National University and the NRF grant funded by the Korean government (MSIT) (grant number 2021R1A2C1011718, RS-2024-00416859). This work was supported by K-GMT Science Program (PID: GS-2023A-Q-207) of Korea Astronomy and Space Science Institute (KASI). ZG is supported by the ANID FONDECYT Postdoctoral program No. 3220029. This work was funded by ANID, Millennium Science Initiative, AIM23-0001.  GJH is supported by the National Key R\&D program of China 2022YFA1603102 and by general grant 12173003 from the National Natural Science Foundation of China.

This research has made use of the NASA/IPAC Infrared Science Archive, which is funded by the National Aeronautics and Space Administration and operated by the California Institute of Technology. This work used the Immersion Grating Infrared Spectrometer (IGRINS) that was developed under a collaboration between the University of Texas at Austin and the Korea Astronomy and Space Science Institute (KASI) with the financial support of the US National Science Foundation 27 under grants AST-1229522 and AST-1702267, of the University of Texas at Austin, and of the Korean GMT Project of KASI. Based on observations obtained at the international Gemini Observatory, a program of NSF NOIRLab, which is managed by the Association of Universities for Research in Astronomy (AURA) under a cooperative agreement with the U.S. National Science Foundation on behalf of the Gemini Observatory partnership: the U.S. National Science Foundation (United States), National Research Council (Canada), Agencia Nacional de Investigaci\'{o}n y Desarrollo (Chile), Ministerio de Ciencia, Tecnolog\'{i}a e Innovaci\'{o}n (Argentina), Minist\'{e}rio da Ci\^{e}ncia, Tecnologia, Inova\c{c}\~{o}es e Comunica\c{c}\~{o}es (Brazil), and Korea Astronomy and Space Science Institute (Republic of Korea).


%

\vspace{1mm}
\facilities{IRTF:Spex, WISE, Gemini:IGRINS}

\appendix

\section{Sources}\label{sec:app1}

In this section we provide some additional information about the SPICY objects with spectroscopic follow-up. Table \ref{tab:app} contains the values for the spectral index and distance found in the literature for the 20 sources.

\movetableright=-0.5in
\begin{table}
	\centering
	\caption{Information on classification and distance for the 20 SPICY YSO candidates with spectroscopic data.}
	\label{tab:app}
\resizebox{\columnwidth}{!}{
   	\begin{tabular}{lccccccc} 
		\hline
 SPICY ID & Other name & $\alpha^{1}$ & \multicolumn{2}{c}{$\alpha^{2}$} &  D$^{3}$ (kpc)&  D$^{4}$ (kpc) &  D$^{5}$ (kpc)\\
\hline
SPICY 11492 & L222\_1 & -0.74 & 0.82 & 0.47 & 1.7 & 9.6 & 10.2 \\
SPICY 15180 & Stim 1 & 0.16 & 0.63 & 0.31 & -- & --  & 10.5\\
SPICY 15470 & 2MASS J13012070$-$6220014 & -1.42  & -- & -- & 3.06 & --  & -- \\
SPICY 21349 & 2MASS J14124874$-$6122507& -0.40 & -- & -- & 3.54  & -- & -- \\
SPICY 29017 & VVVv631 & -0.51 & -0.14 & -- & 1.5 & 2.3 & -- \\
SPICY 31759 & L222\_33 & -0.38 & 0.39 & -0.28 & -- & 3.4  & 4.2 \\
SPICY 35235 & -- & -0.95  & -- & --& -- & -- & -- \\
SPICY 36590 & VVVv270 & 0.13 & 1.8 & 0.26 & -- & 2.3 & 4.6\\
SPICY 42901 & 2MASS J16515774$-$4542390& -0.14 & -- & --& 2.33 & -- & -- \\
SPICY 57130 & 2MASS J17342304$-$3052235& -0.54 &-- & --& --& --& --\\
SPICY 63130 & L222\_148&  1.85& 1.67 & --& 1.5 & 9.1 & 8.3 \\
SPICY 65417 & 2MASS J17482632$-$2407330 & -0.82 & -- & --& 7.39 & -- & -- \\
SPICY 68600 & 2MASS J17550099$-$2801273 & -0.80 & -- & --& 4.29 & -- & -- \\
SPICY 68696 & 2MASS J17551531$-$2852581 & -1.00 & -- & --& 6.96 & -- & -- \\
SPICY 87984 & IRAS 18341$-$0113N & 0.81 & 0.91 & -- & 0.59 & -- &  --\\
SPICY 95397 & SSTGLMC G035.3429-00.4212 & -0.67 & -- & -- & 1.2 & -- &  --\\
SPICY 99341 & 2MASS J19113876$+$0902592 & -0.41 & -0.31  & --& 2.9 & 3 & -- \\
SPICY 100587 &2MASS J19171791$+$1116323& -0.33 & -- & --& -- & -- & -- \\
SPICY 109102 & -- & -0.30 & --& -- & -- & -- & -- \\
SPICY 111302 &-- & -1.44 & -- & -- & 2.2 & -- & -- \\
\hline
\multicolumn{8}{l}{$1$ Spectral index from SPICY}\\
\multicolumn{8}{l}{$2$ Spectral index from the literature. See main text for details.}\\
\multicolumn{8}{l}{$3$ Distance from mean parallaxes of stellar associations found in SPICY.}\\
\multicolumn{8}{l}{$4$ Distance from spatial association with star-forming regions located}\\
\multicolumn{8}{l}{within 5\arcmin ~from the YSO \citep[see][]{2017Contreras_a}}\\
\multicolumn{8}{l}{$5$ Distance from radial velocity measurement of the CO bandhead absorption}\\
	\end{tabular}}
\end{table}

\textit{\textbf{SPICY 11492}}: Designated L222\_1 in \citet{2024Guo_a}, this source is classified as a FUor based on the spectroscopic and photometric characteristics of the YSO.

This is a Class II YSO in SPICY ($\alpha=-0.74$), however, \citet{2024Guo_a} determine $\alpha=0.82$, and \citet{2024Lucas} finds $\alpha=0.47$, which would place it in the Class I YSO category. The difference in classification arises as \citet{2020Kuhn} derives $\alpha$ from the $[4.5]-[8.0]$ colour of the source, while \citet{2024Guo_a} and \citet{2024Lucas} derive the slope from the infrared SED of the source. The difference among the latter arise as \citet{2024Guo_a} include the VVV 2.15 $\mu$m observation, while \citet{2024Lucas} uses wavelengths longer than 3 $>\mu$m, in order to reduce the effect of foreground interstellar extinction.

\citet{2020Kuhn} searched for spatial clustering among their sample of YSOs. SPICY 11492 is found to be associated with group G295.0$-$0.6, which is comprised of 132 YSOs, 31 of which have five-parameter Gaia astrometric solutions. The median parallax of $0.319\pm0.063$ mas yields a distance of d$=1.64^{+0.19}_{-0.15}$ kpc. \citet{2024Guo_a} also provides two values of distance for the source. The first one from possible spatial association with star forming regions ($d_{SFR}=9.6$~kpc) and a second value from radial velocity measurement of the CO bandhead absorption ($d_{RV}=10.2$~kpc).

\textit{\textbf{SPICY 15180}}: This source corresponds to Stim 1 in \citet{2021Guo}, a YSO classified as an emission line eruptive YSO in that work. The authors provide a Class I YSO classification given $\alpha=0.63$. A similar classification is derived from $\alpha=0.31$ estimated by \citet{2024Lucas}. This differs from the flat-spectrum ($\alpha=0.16$) classification given by \citet{2020Kuhn}. The difference arising from using different methods in these works. The only distance information for the source arises from radial velocity measurement, with $d_{RV}=10.5$~kpc.

\textit{\textbf{SPICY 15470}}: This object is a Class II YSO in the SPICY catalog. Prior to this classification, the object was part of the \citet{2008Robitaille} sample of intrinsically red objects from {\it Spitzer}. The object was classified as a candidate AGB star as it showed $[8.0]-[24]<2.5$~mag. 

In the search for spatial clustering, SPICY 15470 is found to be associated with group G304.0$+$02. The latter is comprised of 201 YSOs, 35 of which have five-parameter Gaia astrometric solutions. The median parallax of $0.379\pm0.047$ mas yields a distance of d$=3.06^{+0.51}_{-0.39}$ kpc. 


\textit{\textbf{SPICY 21349}}: Class II YSO in the SPICY catalog, this source has also been classified as a candidate YSO in previous works \citep{2008Robitaille,2016Marton,2019Marton}, but fails to be classified as a pre-MS star in \citet{2020Vioque}.
This source has been observed with {\it Gaia}, where a parallax of $0.84\pm0.42$ mas is presented in the {\it Gaia} DR3 \citep{2023GaiaDR3}. This yields a distance of d$=1.27^{1.45}_{-0.44}$ kpc. The object is also associated with group G311.8$-$0.0 in \citet{2020Kuhn}, a cluster with 1374 members, 209 of which have full {\it Gaia} astrometric solutions. This provides a median parallax of $0.335\pm0.042$ mas, or a distance of d$=3.54^{+0.51}_{-0.46}$ kpc. Given the large uncertainties in the {\it Gaia} parallax of the source, we argue in favor of the distance estimated to the SPICY group G311.8$-$0.0 as the distance of SPICY21349.

\textit{\textbf{SPICY 29017}}: This is VVVv631, an eruptive YSO with an emission line spectrum \citep{2017Contreras,2020Guo}. Different methods yield a classification of this source from Class II ($\alpha=-0.5$) in SPICY to flat-spectrum ($\alpha=-0.14$) in \citet{2017Contreras}. 

\citet{2020Kuhn} find this source to be associated with group G326.6$+$0.6, a cluster with 405 members, 54 of which have full {\it Gaia} astrometric solutions. This provides a median parallax of $0.400\pm 0.045$ mas, or a distance of d$=2.3^{+0.3}_{-0.2}$ kpc. The latter agrees well with the distance to a close SFR provided in \citet{2017Contreras} of d$_{SFR}=2.3$~kpc.

\textit{\textbf{SPICY 31759}}: This is source L222\_33 in \citet{2024Guo_a,2024Lucas}. It was classified as a FUor based on its long duration, high-amplitude ($\Delta$K$_{s}>4.2$~mag) outburst, as well as the characteristics of its near-IR spectrum. This is a Class II YSO in the SPICY catalog ($\alpha=-0.38$), a flat-spectrum source from $\alpha=-0.28$ in \citet{2024Lucas}, while \citet{2024Guo_a} determine $\alpha=0.39$, placing it in the Class I YSO category. Once again, the difference in classification arises from the different methods used to determine $\alpha$

\citet{2024Guo_a} provides a distance of 3.4 kpc (from distance to star forming regions located within 5\arcmin~of the source) and 4.2$\pm$0.5 kpc (from radial velocity measurement of the CO bandhead absorption).

\textit{\textbf{SPICY 35235}}: This is a Class II YSO in the SPICY catalog. There is no previous information in the literature about this source.

\textit{\textbf{SPICY 36590}}: This is source VVVv270 in \citet{2017Contreras_a,2017Contreras,2020Guo}. Classified as an MNor, or eruptive YSO with mixed characteristics between FUors and EX Lupi-type objects, as it shows a long duration outburst, but displays spectra dominated by emission lines. Due to the different methods used to determine a value of the spectral index, $\alpha$, this source is classified as a Flat-Spectrum YSO in SPICY and \citet{2024Lucas}, and as a Class I YSO in \citet{2017Contreras}. From radial velocity measurements, \citet{2017Contreras} determines a distance of 4.6 kpc to the source.

\textit{\textbf{SPICY 42901}}:
Classified as a flat-spectrum source in SPICY, this objects has been previously classified as a YSO candidate in \citet{2008Robitaille} and \citet{2016Marton}. The source is part of the G339.9$-$1.0 group, which contains 121 members with 11 of them having full Gaia astrometric solutions. The mean parallax of these 11 sources provides a distance of d=2.33$^{+1.92}_{-0.72}$~kpc. 


\textit{\textbf{SPICY 57130}}:
This object shares similarities with SPICY 15470 as it is also a Class II YSO in the SPICY catalogue and it was also classified as a candidate AGB star in \citep{2008Robitaille} given that the values of its $[8.0]-[24]$ colour is lower than 2.5 mag. There is no available distance information for this object.

\textit{\textbf{SPICY 63130}}: This is source L222\_148 in \citet{2024Guo_a} where is classified as an emission line eruptive YSO. The spectral index of the source in both \citet{2024Guo_a} and \citet{2020Kuhn}, is consistent with that of a Class I YSO.

The source is part of the G$0.2-0.1$ group, which contains 3662 members with 194 of them having full {\it Gaia} astrometric solutions. The mean parallax of these sources (plx$=0.407\pm0.043$) provides a distance of d$=2.2^{+0.2}_{-0.2}$~kpc. This differs from the distances provided by \citet{2024Guo_a} of d$_{SFR}=9.1$ kpc and d$_{RV}=8.3$ kpc.

\textit{\textbf{SPICY 65417}}: This is a Class II YSO, with no group association in the SPICY catalogue. It was also classified as a candidate YSO in \citet{2019Marton}. The only information in distance arises from \citet{2021Bailer} with d$=7.39^{+2.42}_{-2.06}$ kpc. 

\textit{\textbf{SPICY 68600}}: This Class II YSO from the SPICY catalogue is also classified as  a candidate YSO in \citet{2019Marton}. This source is also part of the analysis of \citet{2020Vioque}, where is classified as a non-YSO. \citet{2008Robitaille} designates this object as a candidate AGB star due to its $[8.0]-[24]$ colour being lower than 2.5 mag. This object is not part of any clusters from the SPICY catalogue. \citet{2021Bailer} provides a distance of d$=4.29^{+4.05}_{-2.14}$ kpc.

\textit{\textbf{SPICY 68696}}: Class II YSO with no group association from the SPICY catalogue. Also classified as a candidate YSO in \citet{2019Marton}. This is also part of the sample of intrinsically ref objects of \citet{2008Robitaille} where is classified as a candidate AGB star based on its $[8.0]-[24]$ colour. \citet{2021Bailer} provides a distance of d$=6.96^{+2.84}_{-2.52}$ kpc.

\textit{\textbf{SPICY 87894}}:  IRAS 18341$-$0113N is part of the sample of \citet{2010Connelley}. It is classified as a Class I YSO in both \citet{2010Connelley} and \citet{2020Kuhn}. 

The source is part of the G29.9$+$2.2 group, which contains 152 members with 36 of them having full {\it Gaia} astrometric solutions. The mean parallax of these sources (plx$=0.384\pm0.074$) provides a distance of d$=2.4^{+0.4}_{-0.3}$~kpc. \citet{2021Bailer} provide a distance of d$=0.59\pm0.06$~kpc. 

\textit{\textbf{SPICY 99341}}: This source was classified as a FUor in \citet{2023Contreras_b} due to the photometric and spectroscopic characteristics of the outburst. This is a Class II YSO ($\alpha=-0.31$) in the SPICY catalogue \citep{2020Kuhn}. \citet{2017Lucas} include it in their sample of high-amplitude variables arising from the UKIDSS GPS survey (source 266, $\Delta$K$=4$~mag). A distance of 3 kpc is estimated due to its possible association with the molecular cloud GRSMC 43.30$-$0.33 \citep{2001Simon}. \citet{2021Bailer} estimate a distance of 2.9$^{+1.2}_{-1.1}$~kpc based on its {\it Gaia} parallax.

\textit{\textbf{SPICY 95397}}: This source is classified as Class II YSO in SPICY. The observation of $^{13}$CO~$\Delta \nu=2$ transitions bandhead absorption, is a clear indicator of the evolved nature of the source. The only distance information comes from the {\it Gaia} parallax of the source, with d$=1.2\pm0.2$ kpc \citep{2021Bailer}.

\textit{\textbf{SPICY 100587}}: Classified as a FUor in \citet{2023Contreras_b}. This is a Class II YSO in SPICY. \citet{2020Kuhn} finds an association with group G45.8-0.4, a cluster with 78 sources. However, only one source has a full {\it Gaia} astrometric solution. Therefore we prefer not to derive a distance based on this information. 

\textit{\textbf{SPICY 109102}}: Classified as a flat-spectrum YSO in SPICY. We classified this source as FUor-like based on its IRTF/Spex spectrum. There is no distance information on this source.

\textit{\textbf{SPICY 111302}}: This source is classified as Class II YSO in SPICY ($\alpha=-1.4$). The observation of $^{13}$CO~$\Delta \nu=2$ bandhead emission is a clear indicator of the evolved nature of the source. The only distance information comes from the {\it Gaia} parallax of the source, with d$=2.2\pm0.3$ kpc \citep{2021Bailer}.

\bibliography{ref_2}{}
\bibliographystyle{aasjournal}



\end{document}